\documentclass[
aps,
prc,
reprint,
twocolumn,
superscriptaddress]
{revtex4-2}

\usepackage{amsmath}
\usepackage{amssymb}
\usepackage{amsthm}
\usepackage{fancyhdr}
\usepackage{graphicx}
\usepackage{hyperref}
\usepackage{multirow}
\usepackage{rotating}
\usepackage{float}
\usepackage{array}
\usepackage{supertabular}
\usepackage{longtable}
\usepackage{amsmath}
\usepackage{dcolumn}
\usepackage{tipa}

\begin{document}

\title{Measuring the cross section of the $^{15}$N($\alpha$,$\gamma$)$^{19}$F reaction \\ using a single-fluid bubble chamber}

% Author List ================================================================
\author{D.~Neto}
\affiliation{Department of Physics, University of Illinois at Chicago, Chicago, Illinois 60607, USA}
\author{K.~Bailey}
\affiliation{Physics Division, Argonne National Laboratory, Lemont, Illinois 60439, USA}
\author{J.~F.~Benesch}
\affiliation{Thomas Jefferson National Accelerator Facility, Newport News, Virginia 23606, USA}
\author{B.~Cade}
\affiliation{Thomas Jefferson National Accelerator Facility, Newport News, Virginia 23606, USA}
\author{B.~DiGiovine}
\altaffiliation{Current address: Los Alamos National Laboratory, Los Alamos, New Mexico 87545, USA}
\affiliation{Physics Division, Argonne National Laboratory, Lemont, Illinois 60439, USA}
\author{A.~Freyberger}
\affiliation{Thomas Jefferson National Accelerator Facility, Newport News, Virginia 23606, USA}
\author{J.~M.~Grames}
\affiliation{Thomas Jefferson National Accelerator Facility, Newport News, Virginia 23606, USA}
\author{A.~Hofler}
\affiliation{Thomas Jefferson National Accelerator Facility, Newport News, Virginia 23606, USA}
\author{R.~J.~Holt}
\affiliation{Kellogg Radiation Laboratory, California Institute of Technology, Pasadena, California 91125, USA}
\author{R.~Kazimi}
\affiliation{Thomas Jefferson National Accelerator Facility, Newport News, Virginia 23606, USA}
\author{D.~Meekins}
\affiliation{Thomas Jefferson National Accelerator Facility, Newport News, Virginia 23606, USA}
\author{M.~McCaughan}
\affiliation{Thomas Jefferson National Accelerator Facility, Newport News, Virginia 23606, USA}
\author{D.~Moser}
\affiliation{Thomas Jefferson National Accelerator Facility, Newport News, Virginia 23606, USA}
\author{T.~O'Connor}
\affiliation{Physics Division, Argonne National Laboratory, Lemont, Illinois 60439, USA}
\author{M.~Poelker}
\affiliation{Thomas Jefferson National Accelerator Facility, Newport News, Virginia 23606, USA}
\author{K.~E.~Rehm}
\affiliation{Physics Division, Argonne National Laboratory, Lemont, Illinois 60439, USA}
\author{S.~Riordan}
\affiliation{Physics Division, Argonne National Laboratory, Lemont, Illinois 60439, USA}
\author{R.~Suleiman}
\affiliation{Thomas Jefferson National Accelerator Facility, Newport News, Virginia 23606, USA}
\author{R.~Talwar}
\affiliation{Physics Division, Argonne National Laboratory, Lemont, Illinois 60439, USA}
\author{C.~Ugalde}
\email{cugalde@uic.edu}
\affiliation{Department of Physics, University of Illinois at Chicago, Chicago, Illinois 60607, USA}

\date{\today}

\begin{abstract}

$^{15}$N($\alpha$,$\gamma$)$^{19}$F is believed to be the primary means of stellar nucleosynthesis of fluorine. Here we present the use of a single-fluid bubble chamber to measure the cross section of the time-inverse photo-dissociation reaction. The method benefits from a luminosity increase of several orders of magnitude due to the use of a thicker liquid target ---when compared to thin films or gas targets--- and from the reciprocity theorem.  
We discuss the results of an experiment at the Thomas Jefferson National Accelerator Facility, where the cross section of the photodisintegration process $^{19}$F($\gamma$, $\alpha$)$^{15}$N was measured by bombarding a superheated fluid of C$_3$F$_8$ with bremsstrahlung $\gamma$-rays produced by impinging a 4 - 5.5 MeV electron beam on a Cu radiator. 

From the photodissociation yield the cross section was extracted by performing a convolution with a Monte Carlo-generated $\gamma$-ray beam spectrum. The measurement produced a cross section that was then time inverted using the reciprocity theorem. The cross section for the $^{15}$N($\alpha$,$\gamma$)$^{19}$F reaction was determined down to a value in the range of hundreds of picobarns. With further improvements of the experimental setup the technique could potentially push cross section measurements down to the single picobarn range.

\end{abstract}

\maketitle

\section{Introduction} \label{sec:introduction}

Radiative capture reactions are of fundamental importance in astrophysics. Protons,
neutrons, and $\alpha$-particles are abundant in many stellar environments and
can interact through ($n,\gamma$),  ($p,\gamma$) and  ($\alpha,\gamma$) reactions
with heavier nuclei under hydrostatic or explosive conditions, or shortly after the Big Bang. 
Reactions involving $\alpha$-particles usually have the lowest cross sections as
the higher Coulomb barrier between the nuclei slows down these capture processes.
In most cases the cross sections are so small that it is difficult to measure these
reactions at stellar conditions in the laboratory using current technologies. For
two of the important astrophysical reactions, $^2$H($\alpha,\gamma$)$^6$Li \cite{Trezzi} and 
$^{12}$C($\alpha,\gamma$)$^{16}$O \cite{deBoer}, the measured cross sections are in the range
of tens of picobarn, thus, requiring a low-background environment, high luminosity and long 
running times.

Most experiments measure the radiative capture cross
sections either in direct kinematics (i.e. a light-ion beam on a heavy target)
or in inverse kinematics (a heavy-ion beam on a light target) usually detecting
the $\gamma$-rays in the exit channel. More recent techniques detect  the recoiling
heavy ion \cite{Schurmann,Fujita} and in more complex experimental setups 
both the $\gamma$-ray and the recoil in coincidence \cite{Ruiz2014}. Ubiquitous 
beam and target contamination and contributions from cosmic rays are usually the main
sources of background that limit the sensitivity of these measurements. Furthermore
the low density of the targets ($\approx$ 1 - 20 \text{$\mu$}g/cm$^2$) prolongs the time needed
to measure the cross sections, thus increasing the contributions from cosmic rays 
and other environmental backgrounds as well.

One possible method for improving the statistics of these measurements is to take
advantage of the time-reversal symmetry of nuclear reactions that involve
strong and electromagnetic interactions and measure the photodisintegration
of nuclei into a light ion (proton, neutron, or $\alpha$-particle) and a heavier
residual nucleus. The cross sections for the reaction X($a$,$b$)Y and the time reversal process 
Y($b$,$a$)X are related via
\begin{equation}
    \frac{\sigma(b,a)}{\sigma(a,b)} = \frac{(2j_X + 1)(2j_a + 1)k^2_{Xa}}{(2j_Y + 1)(2j_b + 1)k^2_{Yb}}
    \label{eq:time_reverse}
\end{equation}
with spins $j$ and wave numbers $k$. Because of phase space transformations, photodisintegration 
reactions can have cross sections which are several orders of magnitude higher
than the corresponding radiative capture processes \cite{BAUR1986188, Ugalde}. Since the
underlying matrix elements are identical for both processes, they can be 
determined by either approach.

The method described in this paper employs the advantages of detailed balance (time-reversal symmetry)
using a thick ($\approx$ 1 - 10 g/cm$^2$) liquid target and
a $\gamma$-ray beam. It can be adapted for measuring some of the most important
nuclear reactions in stellar environments. The luminosity of this technique is
orders of magnitude higher than that of some of the best direct measurements 
performed to date using existing $\gamma$-ray facilities.  

In the experiments discussed below, the residual particles from the photodisintegration
are detected with a bubble chamber \cite{DiGiovine}. The prime example of
a radiative capture process that can be studied with the photodissociation technique
is the $^{12}$C($\alpha$,$\gamma$)$^{16}$O reaction using an oxygen containing
liquid such as N$_2$O. While this reaction is key for understanding the nucleosynthesis
in the universe, it has the complication that oxygen is not a monoisotopic element and, 
thus, requires the use of highly enriched $^{16}$O compounds. 

In a series of experiments we studied the photodisintegration
reaction $^{19}$F($\gamma,\alpha$)$^{15}$N. Since $^{19}$F is a monoisotopic 
element, no background reactions from the photodisintegration of other isotopes
can occur. Since fluorine containing compounds (e.g. CH$_2$FCF$_3$, 
C$_4$F$_{10}$ or C$_3$F$_8$) are used in dark matter experiments \cite{Zacek,Bolte,Ramberg} sufficient information about these liquids in bubble 
chambers is available in the literature. Due to the fact that in the 
$^{15}$N($\alpha,\gamma$)$^{19}$F reaction excited states in $^{19}$F are populated as 
well, no direct comparison between the measured radiative capture and photodissociation 
yields can be made. However, sufficient information about energies, widths and branching
ratios of the critical states in $^{19}$F is available to calculate the expected yields 
for the  $^{19}$F($\gamma,\alpha$)$^{15}$N reaction \cite{Wilmes1,Wilmes2,Dileva}. Filling in Eq. \ref{eq:time_reverse}
for this reaction gives the time-reversal factor as
\begin{equation}
    \frac{\sigma_{(\gamma,\alpha)}}{\sigma_{(\alpha,\gamma)}} = \frac{\mu c^2 E_{CM}}{E^2_{\gamma}}
    \label{eq:ga_to_ag_factor}
\end{equation}
with $\mu$ the reduced mass of the $^{15}$N and $\alpha$-particle, $c$ speed of light, $E_{CM}$ center-of-mass
energy of the $^{15}$N and $\alpha$-particle system, and $E_{\gamma}$ the energy of the resulting
$\gamma$-ray. 

The first set of experiments was performed using a tunable $\gamma$-ray beam from the
HI$\gamma$S facility at Duke University \cite{Weller} produced via inverse
Compton scattering of laser light on electrons circulating in a storage
ring \cite{Ugalde,DiGiovine}.  In these first experiments a good 
agreement between direct ($\alpha$, $\gamma$) measurements and the time-inverse
($\gamma$, $\alpha$) experiments was observed \cite{Ugalde} covering the 
cross section range from 10 \text{$\mu$}b to about 3 nb with the lower cross 
section limit caused by background reactions between electrons and residual 
gas atoms in the storage ring \cite{DiGiovine}. 

In this paper we describe an extension of these measurements towards lower
energies and smaller cross sections using a bremsstrahlung beam from the 
electron injector at Jefferson Lab.

\section{Single-Fluid Bubble Chamber} \label{sec:bubble_chamber}

Bubble chambers were invented more than 60 years ago \cite{Glaser} and have been used as 
detectors for high-energy physics experiments worldwide. During the last decade, they found a
new application as continuously operating superheated detectors in the direct 
search of cold dark matter \cite{DM1,DM2,DM3,DM4}. While the original bubble 
chambers for high-energy experiments are kept in a superheated state for a very short 
time ($\approx$ 1 ms), the dark matter bubble chambers need to be active for extended 
time periods (hours to days). This introduces technical difficulties, as there are 
several processes that can induce bubble nucleation while the detector is waiting for a
signal event. Unwanted bubble nucleation can be avoided by removing the compression 
piston and the buffer fluid or by using a buffer fluid that is in direct contact with 
the superheated fluid. The main difficulty using both an active and a buffer fluid in a 
bubble chamber system originates from chemical reactions and the solubility between the 
active target and the buffer fluid. This can produce solid residues that can be the source 
of unwanted nucleation. For this reason, these two-fluid bubble chambers are 
sometimes referred to as ``dirty" chambers \cite{Waters}.

Single-fluid (or ``clean") bubble chambers have first been used for the
detection of long-lived, low-activity radio-isotopes ($^{14}$C, U or T) using 
diethyl ether or propane \cite{Dodd, Brautti, Waters}. The bubble chamber used here 
employs the same principle. 

\begin{figure}  %figure-1
\includegraphics[width=1.0\columnwidth]{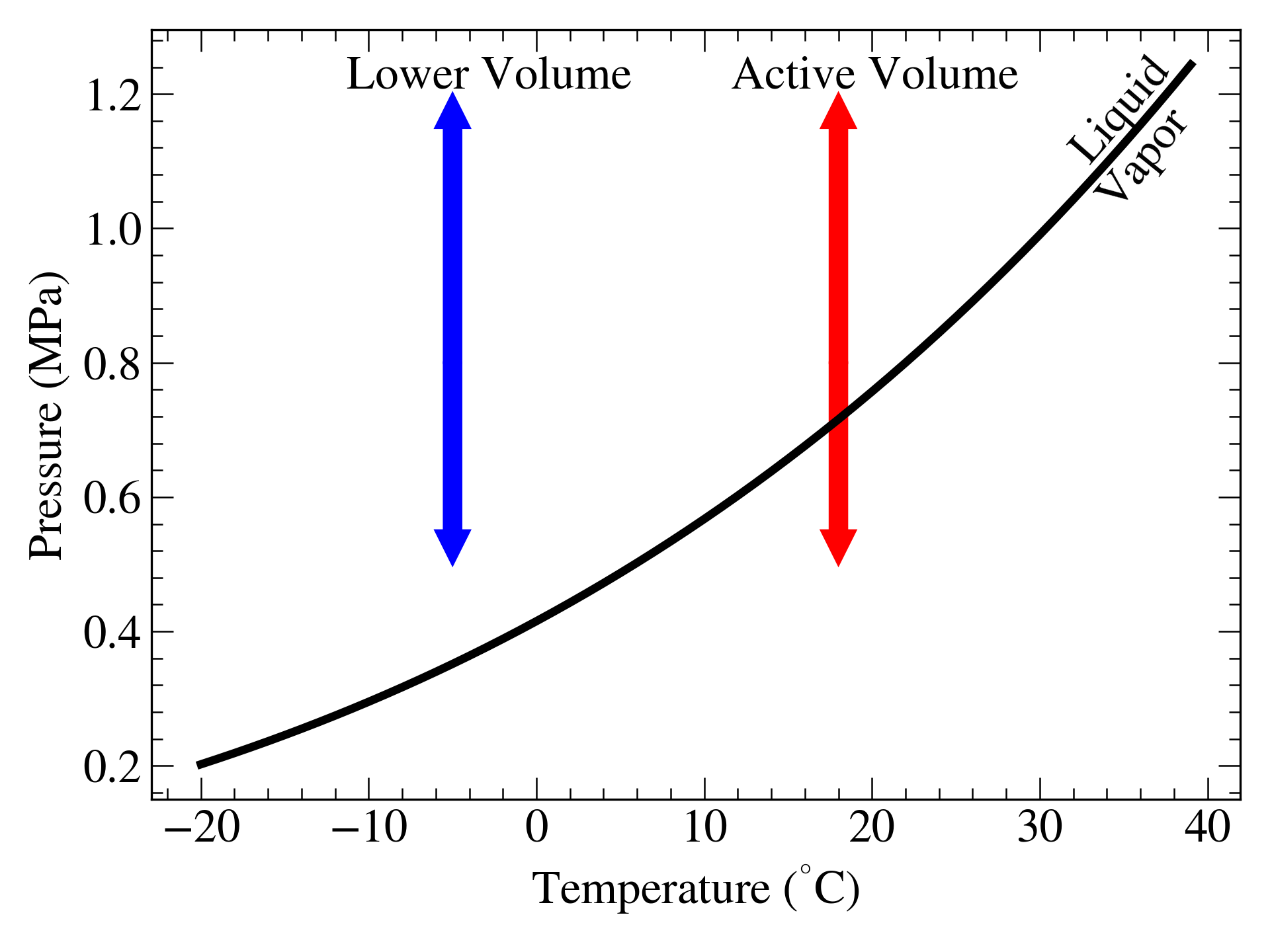} 
    \caption{\label{Fig1}Phase diagram (pressure vs temperature) of C$_3$F$_8$, the active target fluid used in this experiment. The red line shows the region covered in the fiducial area bombarded by the bremsstrahlung beam. This line crosses the phase boundary and creates a superheated liquid which can lead to bubble formation. The blue line represents a cooler region of the glass vessel, such as in the stem of the glass cell. Being at a lower temperature, the liquid does not cross the phase boundary and therefore, does not lead to bubble formation.} 
\end{figure}

\begin{figure}  %figure-2
    \includegraphics[width=1.0\columnwidth]{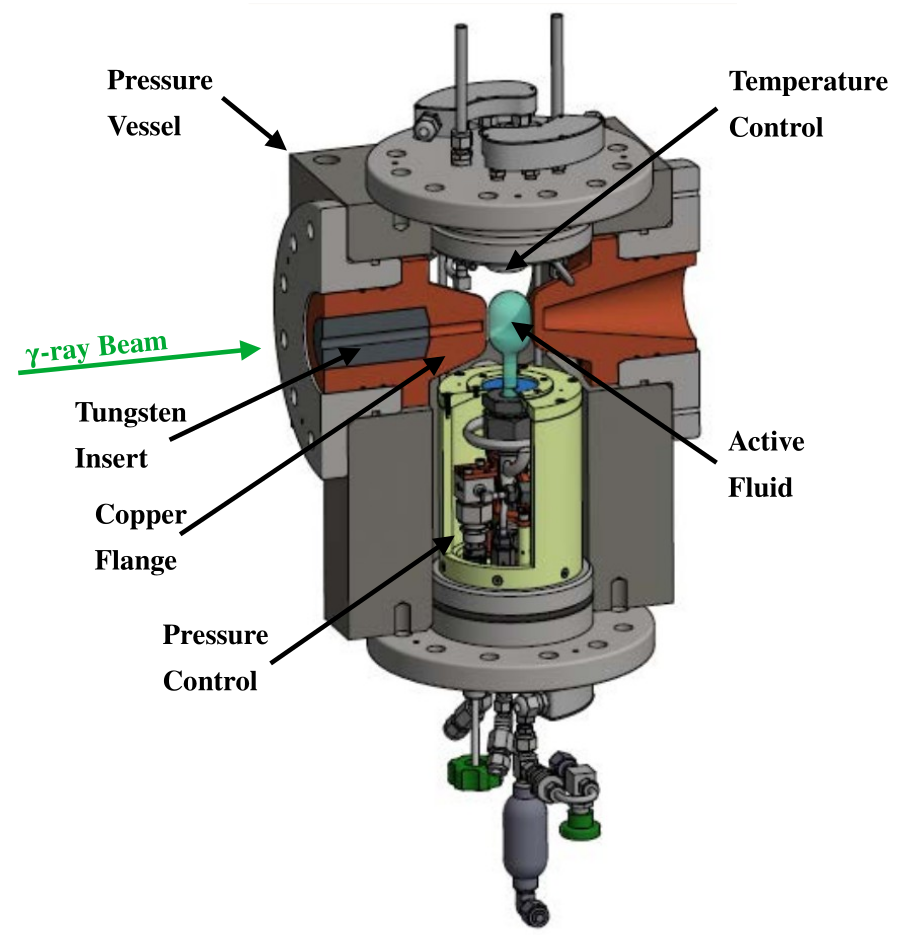} 
    \caption{\label{bubble_chamber_cad}Schematic of the bubble chamber used in the experiment. The $\gamma$-ray beam impinges on the detector from the left. The C$_3$F$_8$ is located in a glass vessel with a long neck that extends down to the region of the hydraulic pressure control system. The temperature is regulated by a mineral oil that surrounds the glass vessel. There is a temperature gradient between the upper region of the glass vessel and the lower region around the glass stem. See text for details.}
\end{figure}

The operational principle of a single-fluid bubble chamber can be seen in  
Fig. \ref{Fig1}, which shows the phase diagram of C$_3$F$_8$ \cite{c3f8_PVdata}. 
At a temperature of approximately 18$^o$C and a pressure of 1.2 MPa, C$_3$F$_8$ is in its liquid form. 
Lowering the pressure to 0.5 MPa (red line in Fig. \ref{Fig1}) brings the liquid into a 
superheated state which, since the products of a photodisintegration reaction deposit 
energy in the liquid, leads to the formation of a bubble \cite{Seitz}. At a temperature of 
-5$^o$C (blue line in Fig. \ref{Fig1}) and pressures $>$ 0.5 MPa, this region of the liquid 
does not cross the liquid-vapor barrier and, thus, will not result in a superheating of the 
fluid.
A schematic of the single-fluid chamber is shown in Fig. \ref{bubble_chamber_cad}. 
A small cylindrical glass vessel, marked as Active Fluid in Fig. \ref{bubble_chamber_cad}, with an inner diameter of 3.6 cm and height of 6.8 cm with a long neck and 
filled with C$_3$F$_8$ ($T$ = 18$^\circ$C, 
$p$ = 0.58 MPa, $\rho$ = 1.35 g/cm$^3$) is located in a box-shaped high-pressure vessel. The glass vessel 
is surrounded by a mineral oil (85.83 $\pm$ 0.13 \% carbon and 14.05 $\pm$ 0.08 \% hydrogen by weight). The pressure in the high-pressure vessel can be adjusted to control
the amount of superheat in the active fluid. The C$_3$F$_8$ filled glass vessel is bombarded by a 
collimated bremsstrahlung beam of 4 - 5.5 MeV $\gamma$-rays from the injector of the 
electron accelerator at Jefferson Lab. The glass vessel is continuously scanned by a 100 Hz
high-sensitivity complementary metal–oxide–semiconductor (CMOS) camera mounted in a lead-shielded container.  
If $\gamma$-rays from the bremsstrahlung beam interact with the fluorine via the 
$^{19}$F($\gamma,\alpha$)$^{15}$N reaction, the $^{15}$N and $\alpha$-particles in the 
outgoing channel are stopped in the C$_3$F$_8$ liquid, which leads to the formation of a 
bubble in the superheated C$_3$F$_8$. If a bubble is observed by the camera, 10 
consecutive frames taken at 10 ms intervals are stored in the computer providing 
information about the location and the motion of the bubble in the fluid. At the same 
time the pressure in the bubble chamber is increased from 0.58 MPa to 2 MPa, which is above the 
critical pressure for C$_3$F$_8$, thus leading to a quenching of the bubble. After a recovery 
time of $\approx$ 10 s the pressure is again decreased to the superheated region at $\approx$ 0.58 MPa.
Details about the thermodynamics of bubble formation or the pressure control 
system used in the experiment can be found in \cite{DiGiovine}. The main
difference between the single-fluid bubble chamber used in this experiment 
and the one described in \cite{DiGiovine} is the absence of a buffer fluid. 
In order to avoid bubble formation in the C$_3$F$_8$ region which is located outside of
the field of view of the CMOS camera, the whole volume below the glass vessel 
containing the superheated C$_3$F$_8$ is surrounded by a separate cylindrical
container kept at lower temperatures (labeled as Pressure Control in Fig. \ref{bubble_chamber_cad}).

\begin{figure} %figure-3
    \centering
    \includegraphics[width=1.0\columnwidth]{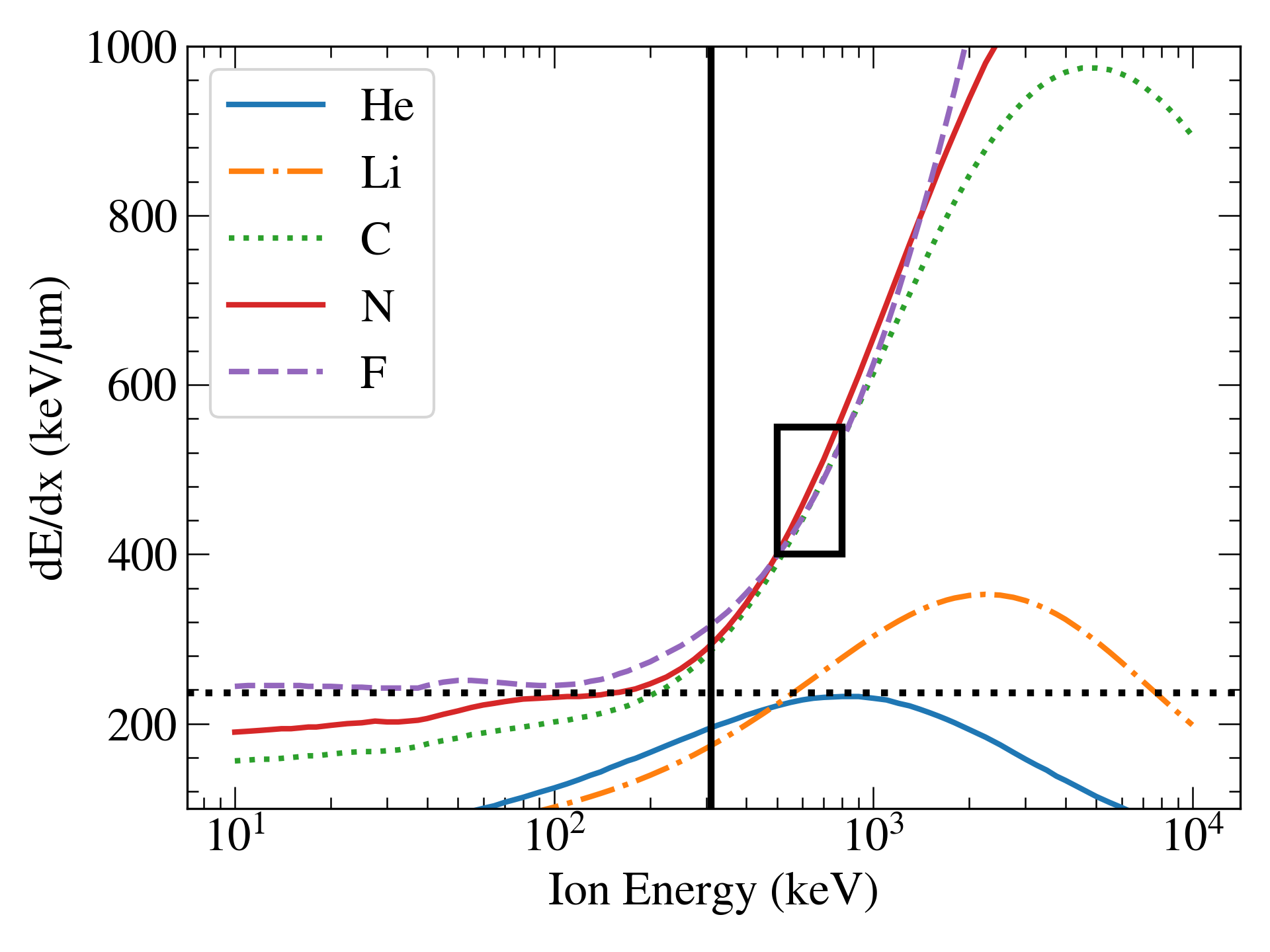}
    \caption{\label{fig:stop_power}Stopping power d$E$/d$x$ vs. energy of various ion species moving in the C$_3$F$_8$ superheated liquid computed using SRIM \cite{ZIEGLER20101818}. The formation of a proto-bubble that will grow into an macroscopic (observable) bubble will depend on the amount of superheat in the liquid. The detector becomes active as the pressure drops. An ion moving into the superheated liquid needs to fulfill two conditions to form a macroscopic bubble: the first condition is to be able to deposit enough energy into the active liquid. The dotted horizontal line represents a model \cite{Harper} for the minimum stopping power (energy deposition per unit length) necessary to induce the formation of an observable bubble. The second condition requires the ion to have enough kinetic energy to form a macroscopic bubble. The vertical solid line is the critical kinetic energy. An ion will trigger the detector when both conditions are met. The region to the right of the vertical solid line and above the dotted line represents the window of detection. The solid line box represents the typical energy range of the recoiling $^{15}$N.}
\end{figure}

The required temperature can be obtained from the $p$-$T$ plot for C$_3$F$_8$ shown in Fig. 
\ref{Fig1}. Operating the bubble chamber at a temperature of 18$^\circ$C in the pressure
range from 0.58 MPa (superheated) to 2 MPa (not superheated) requires a lowering of 
the temperature by about 20 - 25$^\circ$C in the area where bubble generation is to
be prevented. As shown in Fig. \ref{bubble_chamber_cad} a cold region is created using a
cooling circuit inside a cylindrical thermal break (labeled as Pressure Control in Fig. 
\ref{bubble_chamber_cad}) which is kept below -5$^\circ$C. The temperature
distribution was determined with 14 resistance temperature detectors mounted outside the glass vessel. 
Thus, at the temperature of the C$_3$F$_8$ in the lower part of the glass vessel containing the bellows, and the plumbing system, the liquid never crosses the liquid-vapor phase boundary (see Fig. \ref{Fig1}). This allows the same fluid being used as an active target and as a buffer fluid. A plot of the stopping power d$E$/d$x$ vs. energy of various ion species computed in SRIM \cite{ZIEGLER20101818} is shown in Fig. \ref{fig:stop_power}. The full details of bubble formation can be found in Ref. \cite{DiGiovine}.

\section{$\gamma$-ray beam production} \label{sec:accel_details}

\begin{figure*} %figure-4
    \centering
    \includegraphics[width=1.0\textwidth]{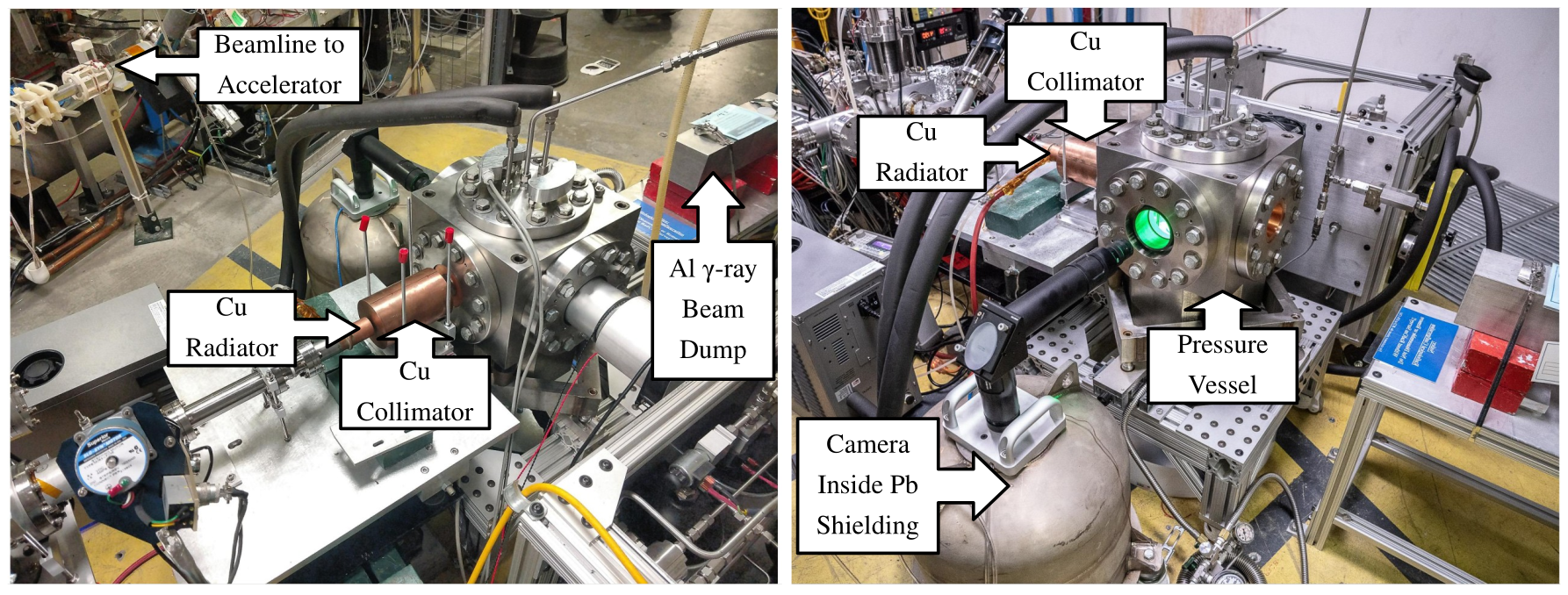}
    \caption{\label{fig:bubble_hall}The bubble chamber experiment in the injector tunnel at Jefferson Lab. The electron beam comes in from the left on the bubble spectrometer line. Down the beamline, the electron beam hits the copper radiator. Bremsstrahlung $\gamma$-rays produced at the radiator enter the pressure vessel through a copper collimator. Additional collimation is provided by a tungsten insert and copper entrance flange (see Fig \ref{bubble_chamber_cad}). Not shown is the additional lead shielding placed around the copper radiator during the experimental runs. Inside the steel pressure vessel (illuminated by a green LED) is the glass cell (which holds the active fluid).}
\end{figure*}

The bremsstrahlung $\gamma$-ray beam was produced by impinging an electron beam
accelerated by Jefferson Lab's injector on a copper radiator (6.0 mm thick, enough to completely stop a 10 MeV/c electron 
beam). The center of the glass cell was located 33 cm away from the radiator face. The injector had
a photo-cathode source operating at 130 kV with
GaAs~\cite{JLABgun} as the photo-cathode material to provide electron beams to
nuclear physics experiments in the experimental halls. After bunching
at 130 keV, the beam was accelerated to 630 keV with a low-Q graded $\beta$ 5-cell
radiofrequency (RF) cavity before being accelerated to relativistic energies (or
nearly relativistic energies as required) in two 5-cell superconducting RF cavities
(quarter cryomodule). Downstream of the quarter cryomodule is a
transport section with three beamlines served by a common dipole: a straight ahead
line to deliver beam to the next stage of acceleration before the beam is
merged into the main accelerator, and two spectrometer dump lines. The bubble 
chamber was installed on one of these two lines (see Fig. \ref{fig:bubble_hall}). Setting and measuring the electron 
beam characteristics for the experiment used all three lines as described in more
detail in Appendix \ref{app:1}. The beam momenta, electron kinetic energy $T_e$ and its associated energy spread, along with the beam vertex at the radiator, are summarized in Table \ref{tab:beam_values}.
\begingroup
\squeezetable
\begin{table}[h]
  \caption{\label{tab:beam_values}Electron beam setups used to measure the yield of the $^{19}$F($\gamma$,$\alpha$)$^{15}$N reaction. An additional measurement was made with an electron beam with kinetic energy of 4.0 MeV, which is sufficiently close to the threshold of $^{19}$F($\gamma$,$\alpha$)$^{15}$N and where the theoretical cross section is orders of magnitude below the beam background limit of this experiment. Since the measurement at $T_e$ = 4.0 MeV does not contribute to the determination of the cross section, in depth analysis was focused on the beam setups with $T_e >$ 4.0 MeV. Electron beam horizontal and vertical vertices are given relative to the center of the copper radiator (in a right-handed coordinate system). A comprehensive study of the electron beam parameters was performed only for the seven setups with $T_e >$ 4.8 MeV, which were used for the measurement of the excitation function discussed. See Appendix \ref{app:1} for details.}
  %\centering
  \begin{ruledtabular}
  \begin{tabular}{ccccc}
  Beam    &  Beam        & Beam      & Beam    & Uncertainty      \\
  Momentum &  Horizontal Vertex & Vertical Vertex & $T_e$ & $\Delta T_e$ \\
  (MeV/c)  &  on Radiator     & on Radiator   & (MeV)  & (MeV)     \\
           &  (mm)            & (mm)          &    &           \\
  \hline
  4.5~~~        &  ~0.2~~                 & -0.3~~                 & 4.0~~~      & 0.1~~~ \\
  5.299        &  ~2.26                 & -1.15                 & 4.813      & 0.010 \\
  5.406        &  ~0.99                 & -5.24                 & 4.919      & 0.010 \\
  5.517        &  -0.29                 & ~0.10                 & 5.030      & 0.010 \\
  5.605        &  -0.78                 & -1.17                 & 5.117      & 0.010 \\
  5.703        &  -0.45                 & ~0.23                 & 5.215      & 0.010 \\
  5.840        &  ~1.02                 & -0.46                 & 5.351      & 0.011 \\
  5.887        &  ~0.95                 & ~0.86                 & 5.398      & 0.011 \\
  \end{tabular}
  \end{ruledtabular}
\end{table}
\endgroup

The electron beam current was measured using a cavity current monitor located
on the main beamline before the common dipole. This current monitor provides
the electron beam current to an accuracy of 3\%.

When the camera detects a bubble,
a signal is sent to the laser table to stop the laser from reaching the photo-cathode for ten seconds.
This allows the bubble chamber to process the bubble, quench the bubble, and restore to the active 
fluid state.

\section {Detection Efficiencies and Background Measurements} \label{sec:detection_and_backgrounds}
\begin{figure*} %figure-5
    \centering
    \includegraphics[width=1.0\textwidth]{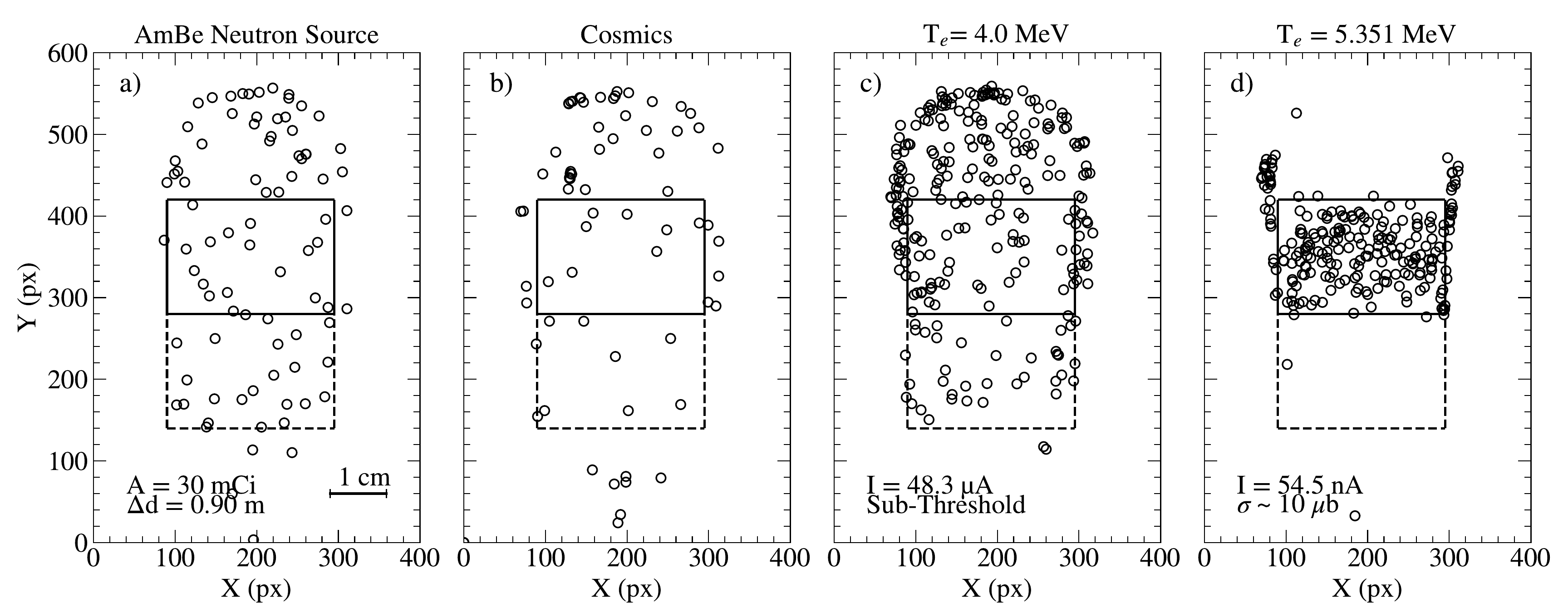}
    \caption{\label{fig:bkg_bubble_dist}Bubble distributions resulting from AmBe neutron source with activity $A$ and at a distance $\Delta d$ from the pressure vessel (a), from cosmic rays (b), and from bremsstrahlung beams produced from electron beams with current $I$ and energies of 4.0 MeV (c), and 5.351 MeV with approximate cross section of the $^{15}$N($\alpha,\gamma$)$^{19}$F reaction for reference (d). The bremsstrahlung beam enters from the left. The vertical and horizontal axes are given in pixels (px), with the position of the bubbles recorded from the CMOS camera. A scale has been added to the first subplot, with 70 px on the CMOS senser corresponding to 1 cm in the plane of the glass cell. At the second highest energy point (d), corresponding to the electron beam energy closest to the strong  $^{19}$F resonance at $E_x$ = 5.337 MeV, the bubbles resulting from the $\gamma$-ray beam form a defined fiducial region (solid black line). This region is used for all energies. Below this fiducial region an equally sized ``background" region (dashed black line) is used for background subtraction. See text for details.}  
\end{figure*}
Since the production efficiency for bubbles depends on the
amount of superheat in the detector \cite{swiss_paper}, the data presented below have been corrected for
changes in pressure and temperature which occurred at the beginning of the experiment. The operating pressure and temperature of the active 
fluid is recorded every ten seconds by the computer control system. From this, we compute the production efficiency for bubbles of the 
particular data collection run and correct the yield accordingly (the uncertainty in the chamber efficiency from the temperature and pressure is $\lesssim$ 3\%). The average value of these corrections per energy are listed in the last column
of Table \ref{tab:summary_table}. 

The response of the single-fluid bubble chamber to incoming neutrons was tested
by exposing the detector to neutrons from a Pu-$^{13}$C source at Argonne National
Laboratory and to an AmBe source while the experiment was setup in the injector tunnel at
Jefferson Lab, with the sources located at distances between 0.9 m and 7 m. The
detection efficiency was found to be homogeneous in the cylindrical section of the
glass vessel, as shown in Fig. \ref{fig:bkg_bubble_dist}a.

There are several possible sources of background events in this experiment. Since bubble 
chambers are insensitive to $\gamma$-radiation, there are no contributions from $\gamma$-ray 
emitting radioactive contaminants such as $^{40}$K. To eliminate contributions from $\alpha$-particle emitting 
isotopes (e.g. Ra, Th, U) which can be present in minute amounts in the material used 
for the construction of the detector, cleaning procedures as described in dark matter
experiments have been employed \cite{RobinsonThesis}, which give typical event rates from the walls 
of the glass vessel of 4 events/day \cite{PhysRevD.93.061101}. 

A second source of background originates from cosmic rays that are detected in the bubble
chamber. The flux of secondary cosmic-ray neutrons at sea level is typically on the order of 0.01 
neutrons/cm$^2$/s \cite{Ziegler}. At the location of the experiment in the injector tunnel, which is $\approx$ 10 
meters water equivalent depth underground, this rate is on the order of 10$^{-4}$ neutrons/cm$^2$/s \cite{GRIEDER2001459}.
Muons, which are after neutrons the second most abundant particles in cosmic rays, 
do not lead to bubble formation under the operating conditions of the bubble chamber used
in this experiment. However, cosmic ray muons can create neutrons via spallation on nuclei.
The cosmic background rate in the active volume has been measured over a period of 76 hrs during the experiment
and was found to be $\approx$ 8$\times$10$^{-3}$ events/s, in good agreement with the expected
flux from cosmic ray induced neutrons enhanced by a minor contribution to the rate from other
sources of natural radiation. A spectrum of these events taken over a period of 2 hours
is shown in panel b) of Fig. \ref{fig:bkg_bubble_dist}. 

A general feature of the bubble distributions shown in the four panels of Fig. \ref{fig:bkg_bubble_dist}
is an increase in the number of bubbles at the interface between the glass and the superheated fluid.
This increase is caused by the presence of boron oxide (typically 15 - 20\%)
in silicate glasses which is added to increase their chemical durability. Incoming low-energy
neutrons from the AmBe source, from cosmic rays or from ($\gamma,n$) neutrons produced in the
material surrounding the bubble chamber can interact with the $^{10}$B in the glass via
the $^{10}$B($n,\alpha$)$^{7}$Li producing $^{4}$He and $^{7}$Li nuclei with energies 
between 1 - 2 MeV, which is sufficient to generate a bubble in the superheated fluid. 
\begin{figure} %figure-6
    \centering
    \includegraphics[width=1.0\columnwidth]{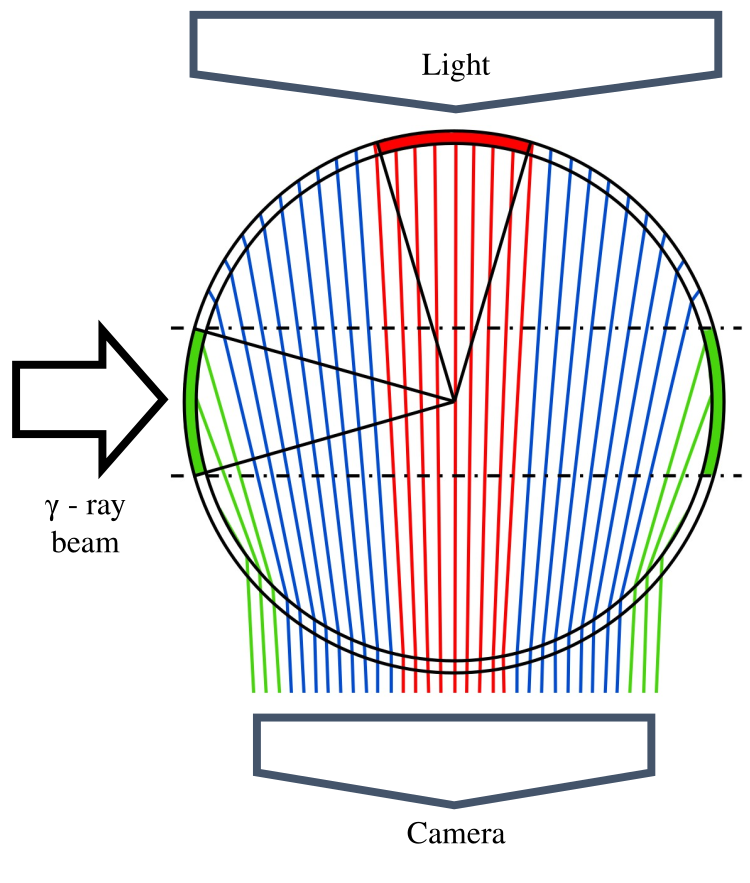}
    \caption{\label{fig:refrac}Effect of refraction on the location of the bubbles from background events  at the glass surface occurring in an angular region covering a range of 30$^{\circ}$ viewed from the top of the glass vessel. Background events from the  $^{10}$B(n,$\alpha$)$^7$Li reaction originating at two regions, along (green) and perpendicular (red) to the incoming beam together with refraction effects lead to an enhancement of the events at the entrance and exit of the $\gamma$-ray beam.}
\end{figure}
The observed spatial distribution of the bubbles is further influenced by refractive effects produced by the glass and liquids. 
Fig. \ref{fig:refrac} shows a simulation of the
refraction effects observed in the bubble chamber consisting of a glass tube ($n$ = 1.47) 
filled with liquid C$_3$F$_8$ ($n$ = 1.22) and surrounded by the mineral oil ($n$ = 1.45) \cite{Lowe1992}. 
In the experiment the glass tube is illuminated from the side and viewed at the 
opposite end by the CMOS camera (see Fig. \ref{fig:bubble_hall}). As can be seen from the calculation the density of events from a section of the 
glass surface located perpendicular to the viewing direction (shown in green) is compressed by a factor 
of $\approx$ 2 when compared to the same area located along the viewing direction (shown in red), leading to a 
concentration of the bubbles on the left and right side of the glass vessel. This effect
is even more pronounced in the dome-like structure at the top of the glass vessel.  Details
will be given in a separate paper \cite{DiGiovine0}. For this reason, the size of the fiducial area determined from a measurment at $T_e$ = 5.351 MeV has been reduced as shown in
Fig. \ref{fig:bkg_bubble_dist}d and the detection efficiency has been corrected accordingly.

\subsection{Non-cosmic backgrounds} \label{subsec:list_beam_bkg}

Candidates for the production of ($\gamma,n$) neutrons from the bremsstrahlung beam interacting 
with materials surrounding the bubble chamber involve isotopes where the neutron 
separation energy is lower than the energy of the incident $\gamma$-rays (e.g. $^{2}$H,
$^{13}$C and $^{17}$O). If the energy of these neutrons is $\gtrsim$ 0.5 MeV they 
can elastically scatter on the superheated C$_3$F$_8$ with the recoiling C and F 
nuclei producing bubble events. As mentioned in the previous paragraph, lower energy neutrons
can produce charged particles through the $^{10}$B($n,\alpha$)$^{7}$Li reaction which
is found to be the main source of the background events. The four primary beam-induced background reactions
are summarized below, with threshold and theoretical abundances from Ref \cite{Audi_2017}.

\begin{itemize}
    \item[a)] $^{2}$H($\gamma,n$)$^{1}$H (threshold = 2.224 MeV, natural abundance = 1.15$\times$10$^{-4}$). Deuterium is present in the mineral oil surrounding the glass cell. Because of its low $Q$-value, the resulting neutrons have sufficient kinetic energy to create bubbles by elastically scattering off the C and F nuclei in the active fluid. Because of its low threshold neutrons from this reaction are present at all energies where measurements were taken.
    \item[b)] $^{13}$C($\gamma,n$)$^{12}$C (threshold = 4.946 MeV, natural abundance = 1.07$\times$10$^{-2}$). $^{13}$C is present in the mineral oil and in the active fluid C$_3$F$_8$. Due to its high reaction threshold, it is only relevant at the highest beam energies. The larger $Q$-value means that most neutrons from this reaction do not have enough kinetic energy to create bubbles by elastic scattering in the active fluid. However, at the highest energies this reaction yields around an order of magnitude higher neutron flux when compared to $^{2}$H($\gamma,n$)$^{1}$H.
    \item[c)] $^{17}$O($\gamma,n$)$^{16}$O (threshold = 4.143 MeV, natural abundance = 3.8$\times$10$^{-4}$). Oxygen is present in the glass and the oil surrounding the superheated fluid. The higher threshold compared to deuterium yields neutrons with insufficient energies to create bubbles via elastic scattering, but oxygen is a  very abundant element in the bubble chamber.
    \item[d)] $^{10}$B($n,\alpha$)$^{7}$Li (threshold = 0.0 MeV, natural abundance = 0.199). $^{10}$B is present in the borosilicate glass. The ($n,\alpha$) reaction occurring at the C$_3$F$_8$-glass interface is the possible source of surface events discussed above.
\end{itemize}

\subsection{Effects of beam-induced background} \label{subsec:effect_beam_bkg}

\begin{figure} %figure-7
    \centering
    \includegraphics[width=1.0\columnwidth]{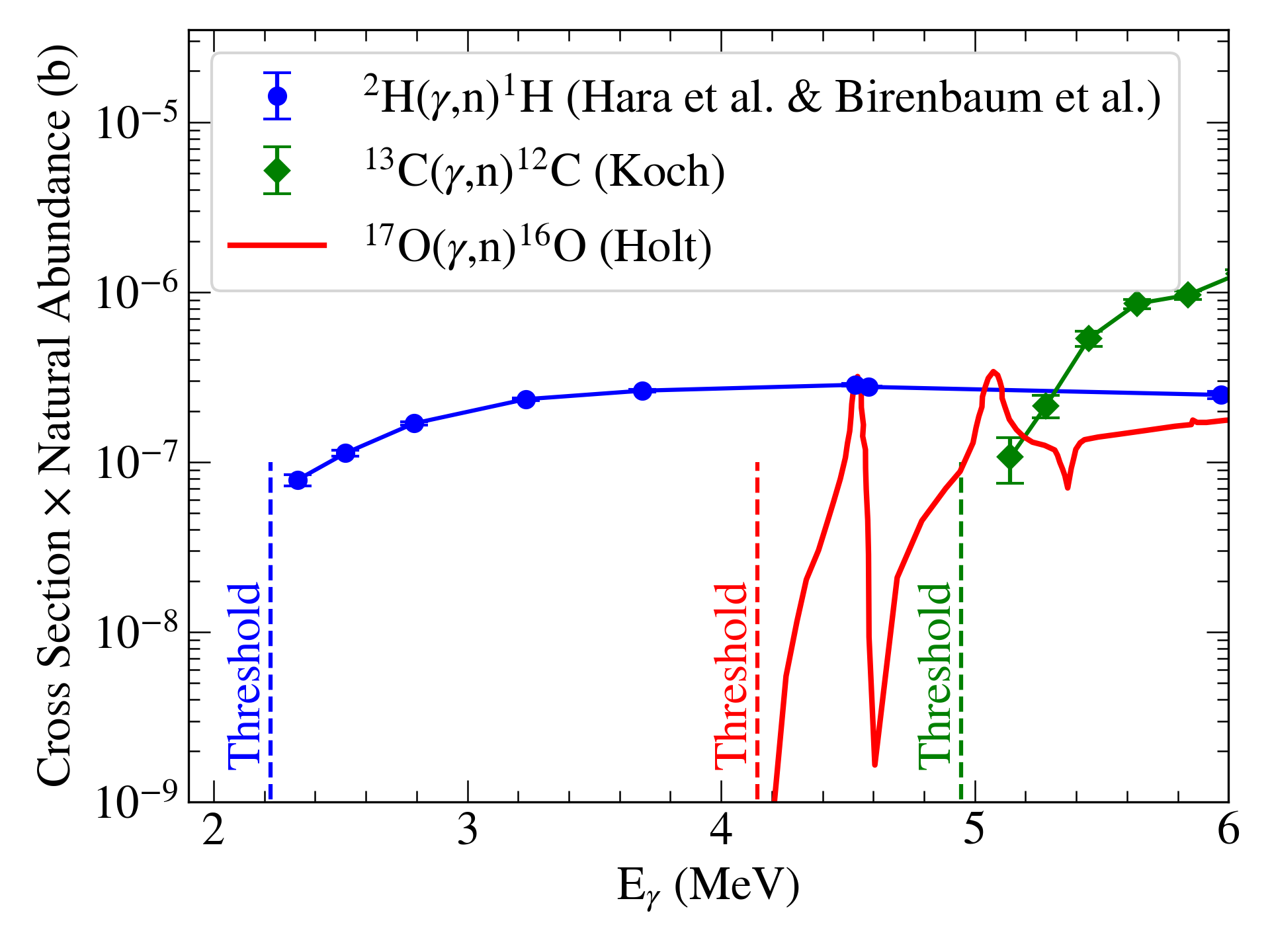}
    \caption{\label{fig:bkg_cs}Cross sections of the three ($\gamma,n$) reactions, multiplied by the natural abundance of each isotope, which can contribute to the beam induced background of the experiment as a function of the $\gamma$-ray energy $E_\gamma$. The threshold of each reaction are denoted by the vertical dashed lines. The solid green and blue curves are interpolations to guide the eye. Over the energy ranges of this experiment, the neutron background predominately comes from $^2$H($\gamma,n$) with $^{13}$C($\gamma,n$) becoming important at higher energies. See text for details.}
\end{figure}
The cross sections for the $(\gamma,n)$ reactions on $^{2}$H \cite{2H_g_n_1,2H_g_n_2}, $^{13}$C \cite{13C_g_n} and $^{17}$O \cite{PhysRevC.18.1962} multiplied by the natural abundance are shown in Fig. \ref{fig:bkg_cs}. From these three reactions only $^{2}$H($\gamma,n$)$^{1}$H and $^{17}$O($\gamma,n$)$^{16}$O have low enough $Q$-values in order to produce neutrons in the full energy range covered in this experiment. As shown in Fig. \ref{bubble_chamber_cad}, oxygen, carbon and hydrogen are present in the beam path of the $\gamma$-rays in the walls of the glass vessel and in the mineral oil surrounding the bubble chamber. While the two isotopes $^{17}$O and $^{2}$H have smaller natural abundances, they can still dominate the background events. (See Fig. \ref{fig:bkg_bubble_dist}.) From the ($\gamma,n$) cross sections in Fig. \ref{fig:bkg_cs} one can see that in the energy region $E_{\gamma}$ $<$ 5 MeV  deuterium is the main source of background events, while for higher $\gamma$-ray energies $^{17}$O (and later $^{13}$C) contribute as well. This background can be eliminated in the future by switching to a fluid that does not contain $^2$H (in place of the mineral oil for thermal control). 

The measurement using the 30 mCi AmBe source (Fig. \ref{fig:bkg_bubble_dist}a) shows events created uniformly over the active 
fluid. The neutrons from this source, having a mean energy on the order of 4 MeV, are able to create bubbles by elastically scattering off carbon and 
fluorine nuclei in the active fluid. A similar distribution is seen in the data from the cosmic rays (second from the left). In both, the AmBe 
and cosmic ray data the surface events, which can be seen in the c) and d) of Fig. \ref{fig:bkg_bubble_dist} where the electron beam is on, are 
not present. This is explained by the relative flux of neutrons from the different sources. In the AmBe and cosmic ray case, the neutron flux is
small, while the average neutron energy is quite high. Contrarily, neutrons produced via the ($\gamma,n$) reactions in the mineral oil are much 
higher in flux while the mean energy is considerably lower. Since the cross section for $^{10}$B($n,\alpha$)$^{7}$Li increases as the neutron 
energy gets smaller, the rate of wall events from the resulting lithium nuclei is considerably larger when the beam is on compared to the rate 
of wall events produced by high-energy neutrons from the AmBe source or from cosmic rays.

\section {Experimental Results} \label{sec:results}

An excitation 
function for the photodisintegration reaction $^{19}$F($\gamma,\alpha$)$^{15}$N
was measured in the energy range from 4.0 MeV to 5.4 MeV. The location of the bubble
in the 10 consecutive pictures mentioned in Sec. \ref{sec:bubble_chamber} were analyzed with a 
software package which allowed the selection of bubbles with similar radii and velocities.
Details of this analysis will be published in a separate paper \cite{DiGiovine0}. 

\subsection{Distribution of bubbles} \label{subsec:bubble_dist}
\begin{figure*}[htb] %figure-8
    \includegraphics[width=\textwidth]{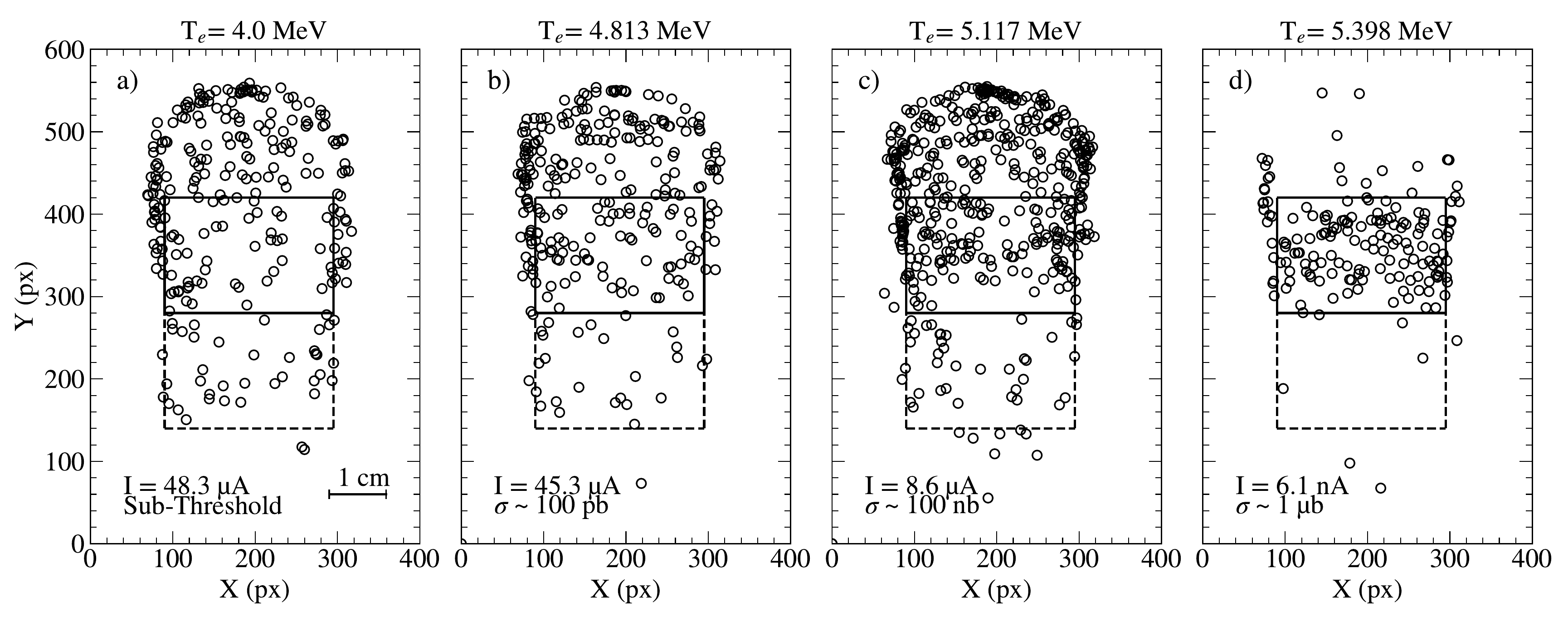}
    \caption{\label{fig:bubbledis}Distribution of events measured in the bubble chamber at various bombarding energies using a bremsstrahlung beam produced from an electron beam with energies of $T_{e}$. The bremsstrahlung beam enters from the left. The electron currents $I$ as well as the estimated cross sections of the $^{15}$N($\alpha,\gamma$)$^{19}$F reaction at these energies are included. The approximate run time of each distribution shown is 1 hr.}
\end{figure*}
The location of bubble events taken at four energies with electron beam currents 
covering the range from around 6 nA to 45 \text{$\mu$}A is shown in Fig. \ref{fig:bubbledis}. At the highest
energies the data overlaps with the previous experiment performed at the HI$\gamma$S
facility \cite{Ugalde} and extends to 4.0 MeV which is below the $^{19}$F($\gamma,\alpha$)$^{15}$N
threshold located at 4.014 MeV. The cross sections calculated from the known
resonance parameters and branching ratios cover the range from 6 \text{$\mu$}b to
100 pb. These cross sections have then to be folded with the energy 
distribution of the bremsstrahlung beam which will be discussed in Section \ref{subsec:cross section}.
Before the bremsstrahlung beam reaches the glass cell it first passes through a 15.24 cm long copper collimator (seen in Fig 
\ref{fig:bubble_hall} with an inner diameter of 0.8 cm and outer diameter of 10.16 cm) then through additional collimation provided by a tungsten insert and copper flange (seen in Fig \ref{bubble_chamber_cad} both with an inner diameter of 1.0 cm).
Thus the photodissociation events in the C$_3$F$_8$ fluid have to be located in 
a cylinder-shaped fiducial area which is shown by the solid lines in Fig. \ref{fig:bkg_bubble_dist} and \ref{fig:bubbledis}. 
Events on the right and left side of the fiducial area are caused by background
events in the wall of the glass vessel (e.g. from the $^{10}$B($n,\alpha$)$^7$Li 
reaction) \cite{DiGiovine0}. In order to subtract
events from cosmic rays, a background area was defined below the fiducial area
as shown by the dot-dashed line in Fig. \ref{fig:bubbledis}. This background area has the same volume as the fiducial area, and since both areas are within the body 
of the cylinder (and do not overlap with the spherical portion of the glass cell) the background region sees the same cosmic rate and effects of diffraction as the 
fiducial region.

\subsection{Experimental yield} \label{subsec:yield}

For each of the experimental runs an experimental yield $Y_{exp}$ is defined by Eq. \ref{eq:yield_eq} 
\begin{equation}
    Y_{exp} = \frac{F - B}{I(t_{tot} - \tau N_{tot})}
    \label{eq:yield_eq}
\end{equation}
where $F$ and $B$ are the number of bubbles observed in the fiducial and background areas, respectively, $N_{tot}$ is the total number of all bubbles detected, $t_{tot}$ the total runtime, $\tau$ the deadtime (10.5 sec) and $I$ is the incident electron beam current  (in \text{$\mu$}A). The yield then amounts to the number of bubbles per deadtime-corrected electron beam charge. The deadtime was determined experimentally as the time it took after the detection of a bubble for the chamber to go from the high pressure state back to the set operating pressure. The range in yield covered in this experiment extends over more than four orders of magnitude. A summary of the experimental values summed over each of the electron beam energies is listed in Table \ref{tab:summary_table}.     
\begin{table*}%[h]
  \caption{\label{tab:summary_table}Summary of all experimental runs at the given electron beam energy listed in the first column. The second and third column are the deadtime corrected charge and runtime, respectively. In the fourth column is the number of $\gamma$-rays $N_{\gamma}$ with kinetic energy above 4.0 MeV which enter the active fluid from the GEANT4 simulation. This number has been scaled such that the charge used in the simulation matches the total charge listed in column two. $N_{tot}$ is the total number of all bubble trigger events for the given energy. From the total active runtime and the measured cosmic rate we determine an estimate for the number of bubble triggers from cosmic rays over the entire active volume. In the next two columns we list the number of events in the fiducial region $F$ and in the background region $B$ corresponding to the solid black lines and dashed black lines in Fig. \ref{fig:bkg_bubble_dist} and \ref{fig:bubbledis}. The final column lists the average correction for bubble chamber efficiency at the given energy. Measurements with the largest bubble efficiency correction factors were done early in the experiment, where some fine-tuning of pressure and temperature was needed.}
  \centering
  \begin{ruledtabular}
    \begin{tabular}{ccccccccc}
    Electron       & Total    & Total    & $N_{\gamma}$($T_{\gamma}$ $\geq$4.0) & $N_{tot}$ & Estimated        & $F$ & $B$ & \multicolumn{1}{c}{Average}  \\
    Beam           & Charge   & Active   & from GEANT4                          &         & Number of        &     & & \multicolumn{1}{c}{Bubble}    \\
    Kinetic Energy & (\text{$\mu$}C) & Run Time & Simulation                           &         & Cosmic Events    &     & & \multicolumn{1}{c}{Efficiency}    \\
    (MeV)          &          & (s)      &                                      &         & in Active Volume &     & & \multicolumn{1}{c}{Correction}   \\
    \hline
    4.0~~~~ & 1.44 $\times10^{5}$ & ~3,042 & - & ~~328  & ~25 & ~57 & ~40 & 1.20 \\
    4.813~ & 4.71 $\times 10^{5}$ & 21,093 & 7.36 $\times 10^{12}$ & 3,472 & 171 & 791 & 419 & 1.35 \\
    4.919\footnotemark[1] & 7.55 $\times 10^{5}$ & 15,964 & 6.37 $\times 10^{12}$ & 2,316 & 130 & 505 & 226 & 1.45 \\
    5.030~ & 9.51 $\times 10^{4}$ & 13,461 & 2.53 $\times 10^{12}$ & 1,715 & 109 & 420 & 209 & 1.15 \\
    5.117~ & 1.84 $\times 10^{5}$ & ~7,811 & 5.79 $\times 10^{12}$ & 1,032 & ~63 & 280 & ~73 & 1.60 \\
    5.215~ & 3.98 $\times 10^{4}$ & 29,400 & 1.52 $\times 10^{12}$ & 1,440 & 239 & 434 & 165 & 1.10 \\
    5.351~ & 1.68 $\times 10^{2}$ & ~6,671 & 8.11 $\times 10^{9}$~ & ~~478 & ~54 & 323  & ~13 & 1.00 \\
    5.398~ & 3.79 $\times 10^{1}$ & ~7,175 & 1.97 $\times 10^{9}$~ & ~~397 & ~58 & 262 &  ~16 & 1.00 \\
    \end{tabular}
  \end{ruledtabular}
  \footnotetext[1]{In the analysis, the vertex of the electron beam at this energy was found to be around 5 mm below the center of the copper radiator. As a result, the number of $\gamma$-rays reaching the glass cell was smaller by about a factor of 2.8 in comparison to adjacent beam energies (when correcting for beam current and energy). The yield resulting at this electron energy has been corrected accordingly and the uncertainties have been increased by the same factor.}
\end{table*}

\begin{figure} %figure-9
    \centering
    \includegraphics[width=1.0\columnwidth]{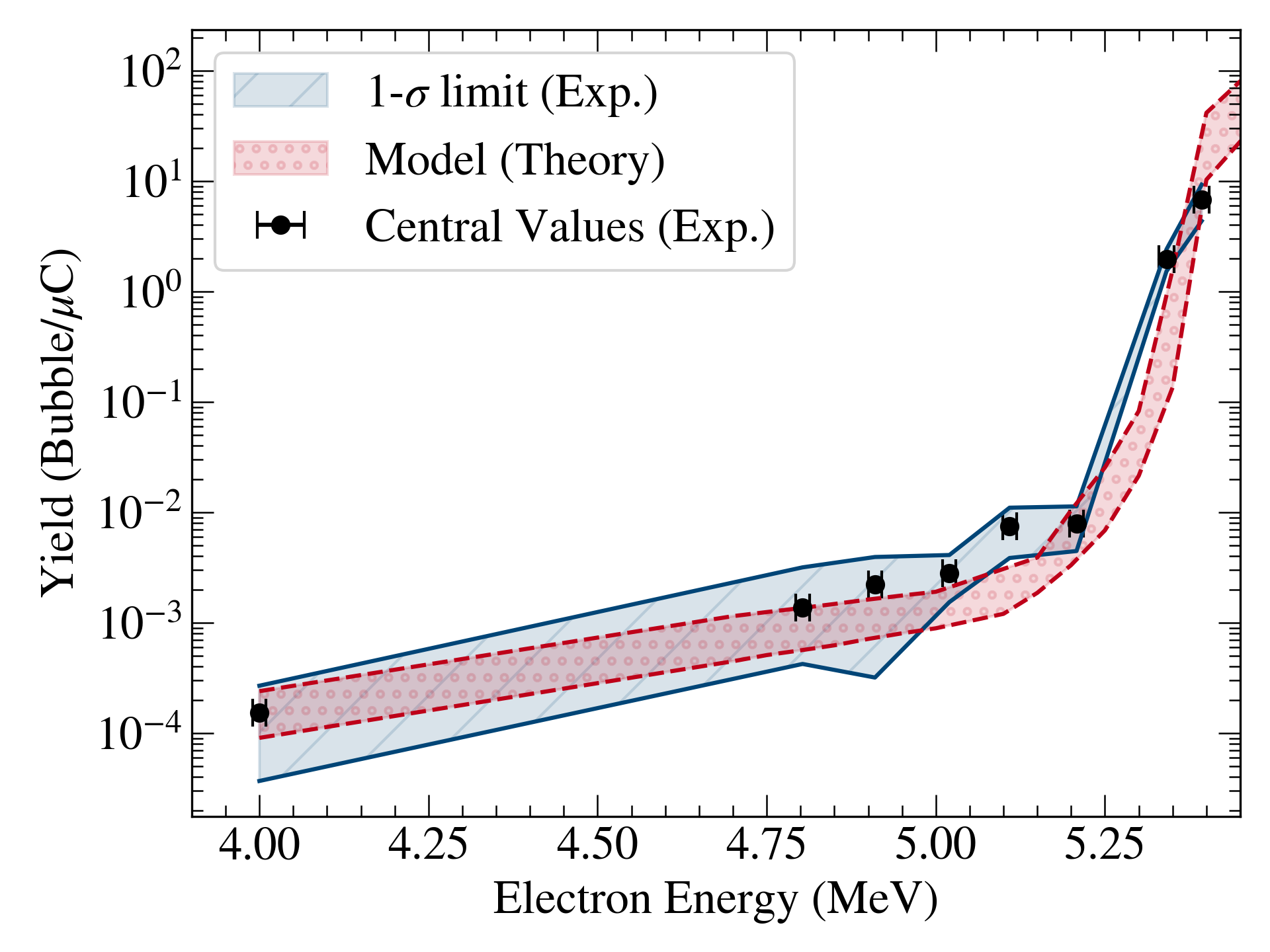}
    \caption{\label{fig:yield_fig}Experimental yield defined by the Eq. \ref{eq:yield_eq}. The black dots represent the central values from the weighted averaging with horizontal bars for the measured spread in the electron beam energy. The hatched area is the 68\% confidence limit from the MC calculation. The interpolation of the hatched fill along with the upper and lower lines which bound the fill is only to guide the eye.  The dashed red lines are the theoretical yield curve from a model (described in Appendix \ref{app:2}) which includes the effect of the dominant background resulting from $^{2}$H($\gamma,n$)$^{1}$H.}
\end{figure}

For each energy, a weighted average yield is computed which gives the central values (black dots) in Fig. \ref{fig:yield_fig}. The statistical weight of each run, within a given energy, is set by the quantity $F - B$. To determine confidence intervals, a simple Monte Carlo (MC) calculation samples over the parameters of Eq. \ref{eq:yield_eq}. The beam current for each run is recorded by a current monitor (see details in Sec. \ref{sec:accel_details}), from which the average beam current with Gaussian error bars for each run is determined.  For $F$, $B$ and $N_{tot}$ error bars are assumed from simple counting errors. From the MC sampling the Gaussian 68\% error bars are shown as the hatched region in Fig. \ref{fig:yield_fig}. In Fig. \ref{fig:yield_fig} we also show a curve bound by two red dashed lines which is a theoretical yield curve from a model as described in App. \ref{app:2}. 

\subsection{GEANT4 simulations} \label{subsec:GEANT4_sim}

From the measured electron beam parameters, the resulting bremsstrahlung beam 
is determined using GEANT4 \cite{Geant1, Geant2, Geant3}. Surveys taken at 
Jefferson Lab prior to the experiment recorded the relative positions of the copper 
radiator, copper 
collimator, pressure vessel,
and aluminium beam dump. From this survey the components
in the simulation are aligned to match the experimental conditions.

The choice of physics list for the simulation is dictated by the need for
an accurate simulation of the production of the bremsstahlung beam from the copper
radiator, and the transport of the $\gamma$-rays through to the glass cell. For the 
electromagnetic physics, the GEANT4 Livermore E\&M library (EM Liv) was used. 
This uses the Seltzer-Berger model \cite{SELTZER198595} at the energy range of 
this experiment. For the hadronic physics three models were tested: a Bertini cascade model \cite{GUTHRIE196829, Heikkinen:2003sc}, 
a Bertini cascasde with high precision neutron data \cite{NeutronHP} (ENDF/B-VIII.0), and the Binary cascade model \cite{Folger2004}. Since the complete characterization of the backgrounds using GEANT4
was outside the scope of this work (but will be performed in a future work), the simulation was focused on producing a
high quality $\gamma$-ray spectrum inside the glass cell. Little changes
were seen across the three different hadronic physics models. The Bertini cascade model without high precision neutron data
was selected for the final production simulations as it performed slightly faster. 

To check the physics and geometry of the GEANT4 model, simple simulations were 
performed to model the AmBe neutron source tests with the number of neutrons
and their energy spectrum recorded for simulations of the source at the same three
distances tested with the bubble experiment at the Jefferson Lab injector tunnel.
From these neutron spectra the estimated number of bubbles in the active fluid
was found to be in good agreement with the experimental results at the two largest
distances, with the result of the simulation of the smallest distance being
within the 2$\sigma$ error bars of the experimental result.
\begin{figure}%[htb] %figure-10
    \includegraphics[width=\columnwidth]{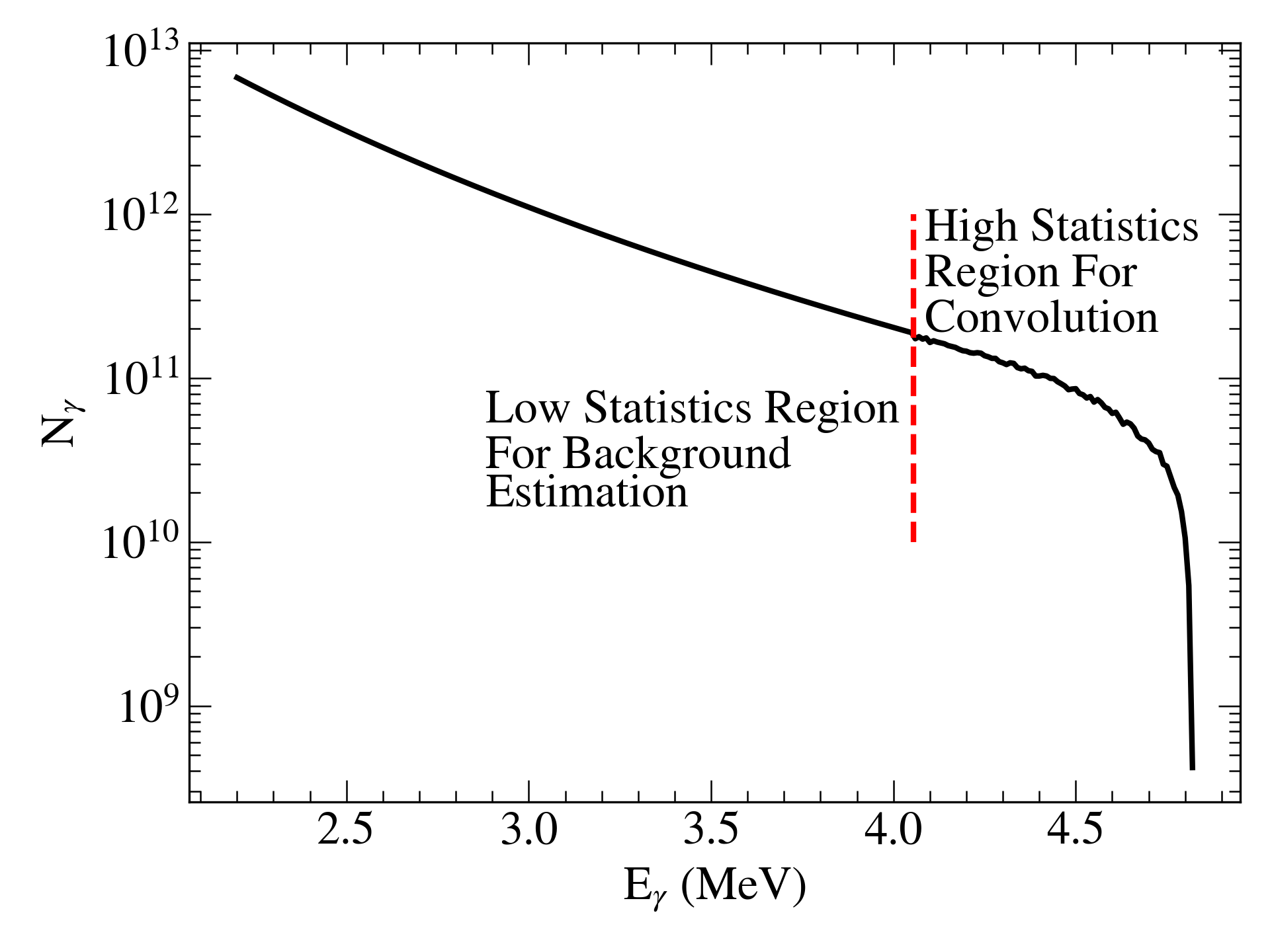}
    \caption{\label{fig:geant4_brem_spectra}Bremsstrahlung spectra inside the glass cell from the GEANT4 simulation of the electron beam at $T_e$ = 4.813 MeV. The dashed vertical line marks where the two simulations described in Sec. \ref{subsec:GEANT4_sim} are stitched together. The high statistics region resulting from a simulation of $10^{11}$ electrons using an energy cut at 3.8 MeV. The low statistics region resulting from a simulation of $10^{9}$ electrons using an energy cut at 2.2 MeV. The lower statistics data set is scaled and combined to the high statistics set near the $^{19}$F($\gamma$,$\alpha$)$^{15}$N reaction threshold (matched over a range extending around 100 keV to the left of the red dashed vertical line). The result shown here is then scaled to match the total beam charge of the $T_e$ = 4.813 MeV experimental run listed in Table. \ref{tab:summary_table}.}
\end{figure}

GEANT4 simulations were performed at each of the principal energies listed in
Tab. \ref{tab:beam_values} in addition to a simulation using a beam with an electron kinetic energy of 4.0 MeV. 
For each of these energies two sets of simulations were performed. One set uses $10^{11}$ electrons with an energy cut set 1 MeV below the simulated electron
beam energy to provide high statistics data for the convolution used to determine the 
cross section described in Sec. \ref{subsec:cross section}. A second set of simulations was performed
using $10^{9}$ electrons with an energy cut set at 2.2 MeV. This lower statistics data set is then scaled up and used to fill in the low energy
tail of the higher statistics data set for background estimation (data sets were stitched together over an overlapping range of 100 keV near
the energy cut of the high statistics simulation data). An example of the resulting 
bremsstrahlung spectra can be seen in Fig. \ref{fig:geant4_brem_spectra}. 

In the electron beam analysis, the vertex of the electron beam for the experimental run at $T_e$ = 4.919 MeV was found to be around 5
mm below the center of the copper radiator. GEANT4 simulations of this energy setup showed a decrease in the number of $\gamma$-rays reaching the glass cell by a 
factor of 2.8.  The yield resulting from this electron energy has been corrected for this shift and the 
uncertainties of the yield at $T_e$ = 4.919 MeV have been increased by the same factor.

In the GEANT4 simulations the $\gamma$-rays were tracked in the glass cell (which houses the active fluid).
In addition $\gamma$-rays were tracked directly after the copper radiator to measure the $\gamma$-ray flux in
the forward direction. This allows for a double check of the simulation (electrons with a Gaussian energy
spread hitting a copper radiator is a well known problem) and to check how well the copper collimator and
tungsten alloy insert produce a defined $\gamma$-ray cone in the active fluid. The number of $\gamma$-rays, 
their kinetic energy, and their 2D position on a plane perpendicular to the primary beam axis were recorded. 
From this, the bremsstrahlung profile and the $\gamma$-ray beam shape in the active fluid was constructed.

\begin{figure} %figure-11
    \centering
    \includegraphics[width=1.0\columnwidth]{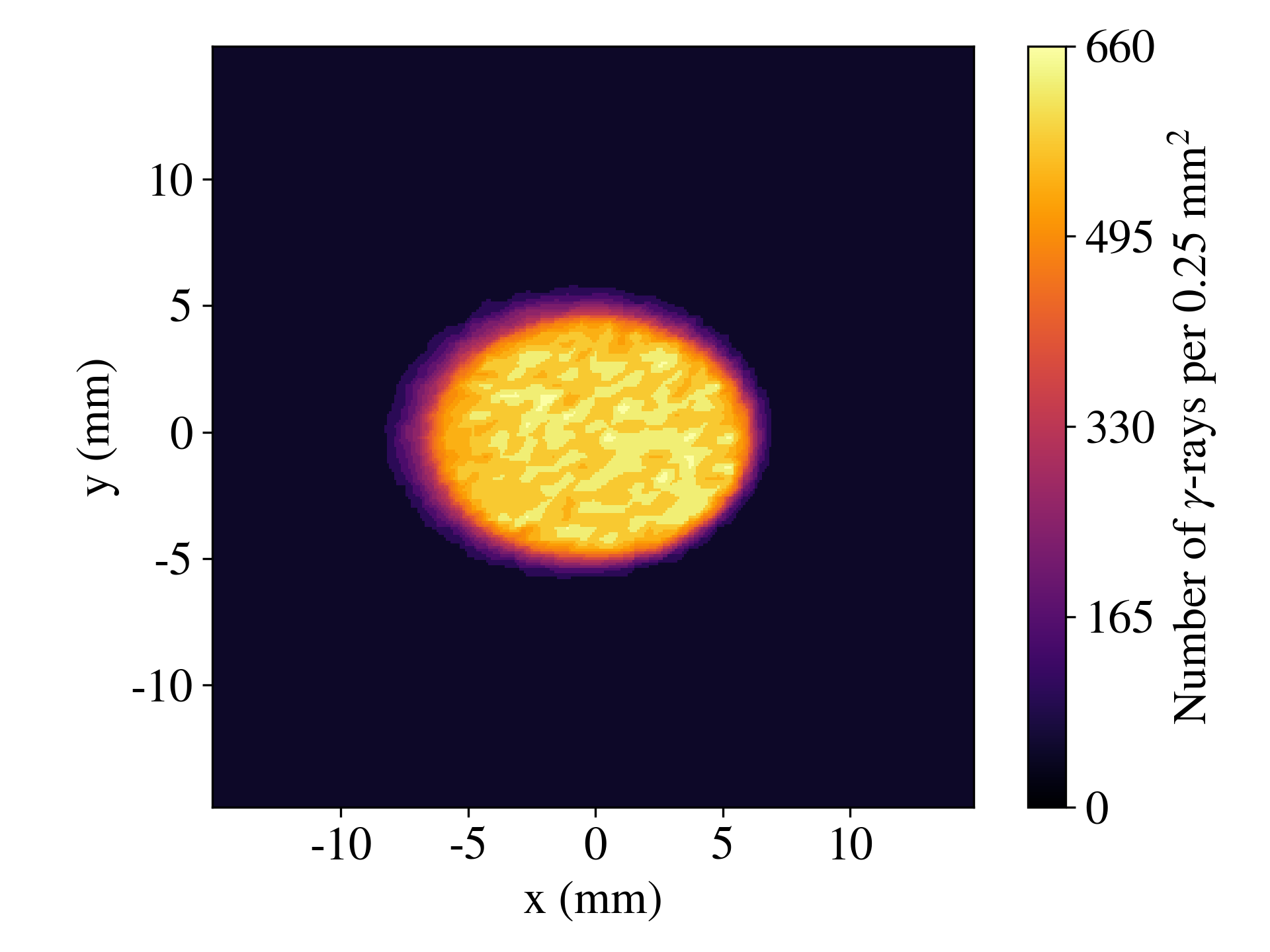}
    \caption{\label{fig:bubble2d} A 2D $\gamma$-ray distribution inside the glass cell from a GEANT4 simulation of the $T_e$ = 4.813 MeV electron beam using 10$^9$ electrons. The origin has been centered relative to the center of the copper radiator. The vertical walls of the glass cell are outside the range of the figure (in this coordinate system, having an $x$ position at approximately +18 and -18 mm).}
\end{figure}

Integrating over the 2D $\gamma$-ray distribution inside the glass cell, a circular area
containing approximately 95\% of the $\gamma$-rays was found to have a diameter
that matched the size of the fiducial region described in Sec. \ref{subsec:bubble_dist}. This 
 diameter is consistent with the height of the solid box in Figs. \ref{fig:bkg_bubble_dist}
and \ref{fig:bubbledis}. An example of the 2D $\gamma$-ray distribution from a GEANT4 simulation
of the $T_e$ = 4.813 MeV electron beam can be seen in Fig. \ref{fig:bubble2d}. The ratio of the number of simulated electrons to the number
of $\gamma$-rays reaching the glass cell was used to scale up the simulation
to match the total beam current at each of the eight beam energies (values in Table \ref{tab:beam_values} 
in addition to the simulation of an electron beam with kinetic energy of 4.0 MeV).

\subsection{Determining the cross section} \label{subsec:cross section}

The experimental yield shown in Fig. \ref{fig:yield_fig} can be recast using the modeled $\gamma$-ray spectrum
detailed in the previous section. This provides a relationship between the number of bubbles estimated to come
from the $^{19}$F($\gamma$,$\alpha$)$^{15}$N reaction as a result of the number of $\gamma$-rays used at a given energy setup.
The relationship between the cross section and theoretical yield is given by a convolution \cite{rolfs_rodney_2005} as 
\begin{equation}
    Y_{th}(E) = \int^{E}_{E_{t}} n_t \tilde{N}_{\gamma}(E,k)\sigma(k)dk
    \label{eq:cs_eq}
\end{equation}
with $Y_{th}(E)$ the theoretical yield at energy $E$, $n_t$ a factor which describes the target thickness in terms of target nuclei per cm$^2$ (for this experimental 
setup $n_t = 9.5 (0.1) \times10^{21}$ $^{19}$F/cm$^2$), $E_{t}$ the threshold energy, $\tilde{N}_{\gamma}(E,k)$ the number of $\gamma$-rays per unit energy --- itself 
a function of 
both the electron and $\gamma$-ray energy --- and $\sigma(k)$ the cross section to be convoluted with the bremsstrahlung spectrum.

To obtain the cross section one can solve this equation in the ``backwards direction" (deconvoluting
the experimental yield with the bremsstrahlung profiles to get the cross section) or 
in the ``forward direction" (convolute the bremsstrahlung profiles using a known or modeled
cross section). Historically, the Penfold-Leiss method \cite{Penfold-Leiss} of deconvolution or unfolding has been the preferred technique. However, this method has
shortcomings that required further improvement \cite{FINDLAY1983353}. Electron beam energy loss and 
the effect of the thickness of the radiator used to produce the bremsstrahlung beam are 
considered in the updated technique. This deconvolution method is also sensitive to the 
size of the error bars of the data ---this becomes more evident if the yield is very flat. The technique also requires a reasonably uniform spread in the data points (typically every 100 keV or so), and special care has to be taken when employing the method near a resonance. 

In general, convolution is much simpler than deconvolution. The process of performing the convolution via an iterative process, 
comparing models to experimental data and using parameters tuned with the help of machine learning methods is being 
developed for future experiments with the bubble chamber. For this work, the bremsstrahlung profiles described in Sec. 
\ref{subsec:GEANT4_sim} were convoluted with a cross section for the $^{19}$F($\gamma,\alpha$)$^{15}$N reaction constructed from the Breit-Wigner model (Eq. \ref{eq:BW}) using resonance parameters from Wilmes et al. \cite{Wilmes2} (see Appendix \ref{app:2} for details). Over the energy range of this work, the cross section is dominated by the resonance at $E_x$ = 5.337 MeV. A recent measurement of the elastic scattering reaction $^{15}$N($\alpha$,$\alpha$)$^{15}$N by Volya et al. \cite{Volya} finds parameters for this resonance in good agreement with the earlier Wilmes et al. results. 

\begin{figure} %figure-12
    \centering
    \includegraphics[width=1.0\columnwidth]{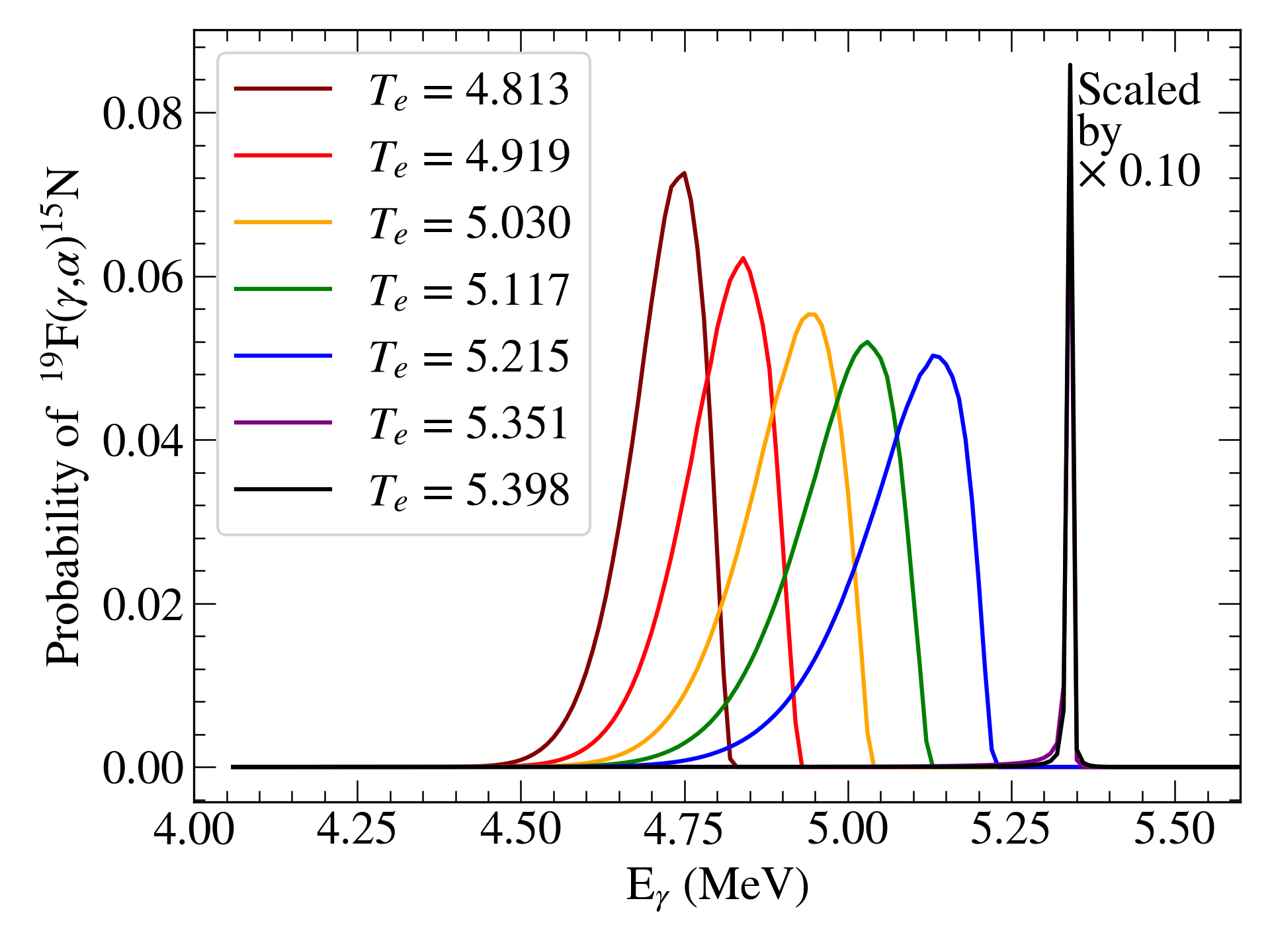}
    \caption{\label{fig:conv_resp_fig}Probability density function (PDF) for bremsstrahlung $\gamma$-rays resulting from an electron beam with the energies listed in Table \ref{tab:beam_values} to photodisintegrate $^{19}$F, based on the calculated cross section described in Appendix \ref{app:2}. The PDF for $T_e$ = 5.351 and 5.398 MeV have been scaled down by a factor of 10. From the peak of each PDF one determines the excitation energy at which the corresponding value of the cross section was measured. The FWHM of each PDF are listed in Table \ref{tab:cs_values}. One can see that even with the broad bremsstrahlung spectrum used, the cross section falls off sufficiently fast at low energies. The energy range of the $\gamma$-rays which contribute to the yield resembles the shape of a peak.}
\end{figure}

From the convolution one can also determine the probability distribution function (PDF) that a $\gamma$-ray with a given energy will lead to a 
disintegration of $^{19}$F in the active fluid. In Fig. \ref{fig:conv_resp_fig} the PDF's resulting from the different electron 
beam energies are shown, with the curves for $T_e$ = 5.351 and 5.398 MeV being scaled down by a factor of 10 to allow for a comparison of the shape 
of all the curves more easily.  The center of the energy bin which contains the peak of the PDF gives the excitation energy of $^{19}$F where the 
cross section for the $^{19}$F($\gamma$,$\alpha$)$^{15}$N reaction was measured. The uncertainty for the excitation energies listed in Table 
\ref{tab:cs_values} are obtained by summing in quadrature the associated $\Delta T_e$ for the electron beam (from Table \ref{tab:beam_values}) with
the step size used in the discrete convolution to compute the PDF (10 keV). The FWHM of the PDF gives the effective size of the $\gamma$-ray energy 
bin.

An additional convolution was performed using the GEANT4 bremsstrahlung profiles and the cross section of the $^{2}$H($\gamma,n$)$^{1}$H 
reaction. This is used in the construction of the 
theoretical model (see Appendix. \ref{app:2} for details) yield curve shown as the dashed black line in Fig. \ref{fig:yield_fig} and for 
modeling beam background for subtraction. 

\begin{figure} %figure-13
    \centering
    \includegraphics[width=1.0\columnwidth]{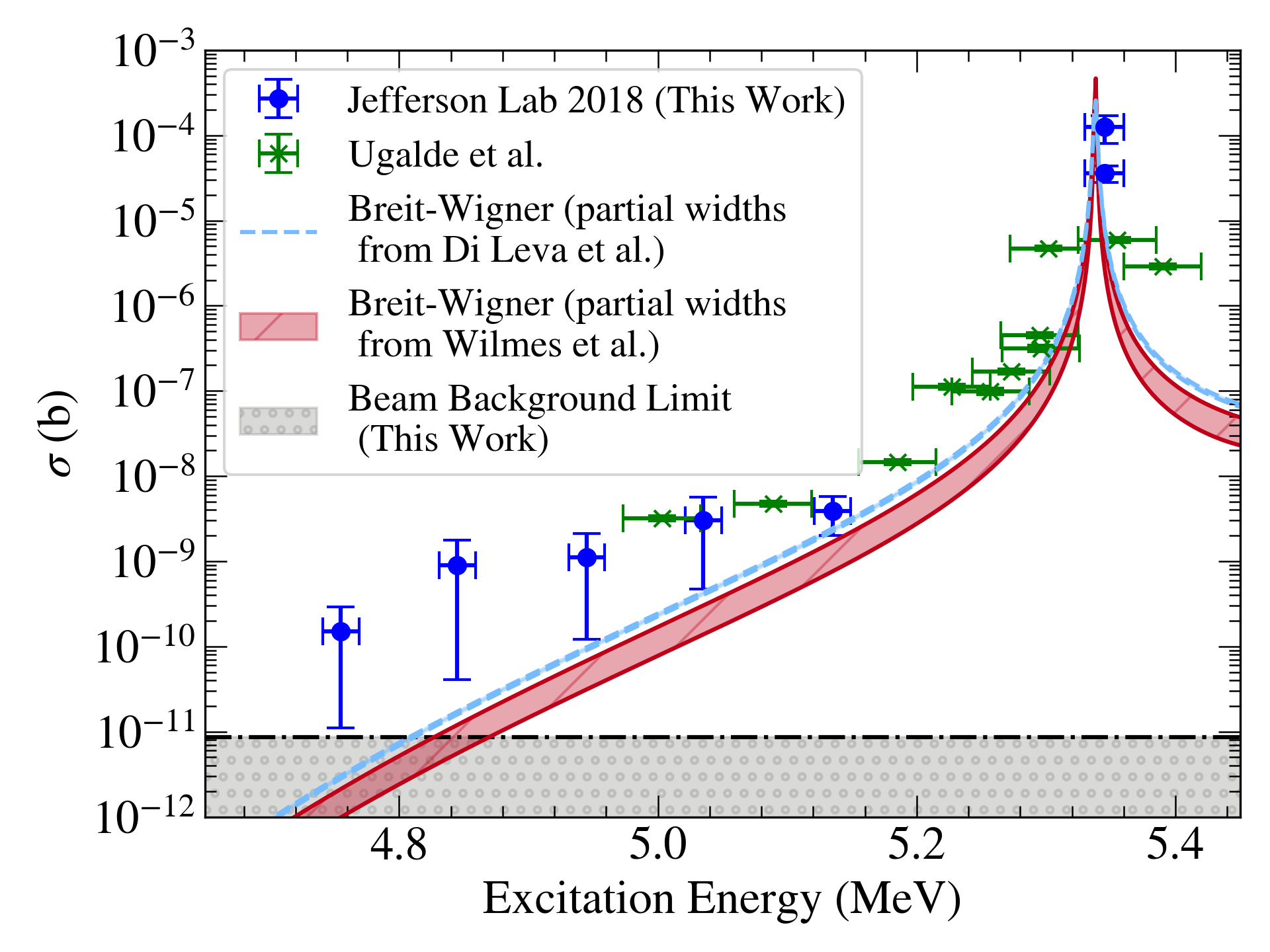}
    \caption{\label{fig:cs_fig}Comparison of experimental cross sections taken from Fig. 12 of Ugalde et al. \cite{Ugalde} as green crosses and this work as blue circular dots with the theoretical Breit-Wigner cross section using partial widths from Wilmes et al. \cite{Wilmes2} as solid red curve and partial widths from Di Leva et al. \cite{Dileva} as dashed blue curve for the $^{15}$N($\alpha$,$\gamma$)$^{19}$F reaction. The hatch fill of the Breit-Wigner curves comes from varying the $\alpha$-particle width of the $E_R$ = 5.337 MeV resonance by the 1$\sigma$ limits measured in each respective paper. The horizontal black dashdot line represents the 1$\sigma$ limits on the beam background of the current bubble chamber setup.}
\end{figure}

From this theoretical model yield we compare to the experimentally measured yield and determine the corresponding 
experimental values of the cross section. Using the time reversal factors between the ($\gamma$,$\alpha$) and ($\alpha$,$\gamma$) 
reactions (the factors listed in second column of Table \ref{tab:cs_values}), we then extract the cross section for the inverse reaction $^{15}$N($\alpha$,$\gamma$)$^{19}$F 
(shown as the blue dots in Fig. \ref{fig:cs_fig}). The inferred cross section of the reaction $^{15}$N($\alpha$,$\gamma$)$^{19}$F, 
along with the corresponding excitation energies, are listed in Table \ref{tab:cs_values}.  From the sub-threshold measurement at 4.0 MeV we 
determined the beam background limit of the current bubble chamber setup represented by the dashdot horizontal line in Fig. 
\ref{fig:cs_fig}.
\begingroup
\squeezetable
\begin{table}[h]
  \caption{\label{tab:cs_values}Summary of the cross section information for the $^{15}$N($\alpha$,$\gamma$)$^{19}$F reaction determined from the experimentally measured yield. Excitation energies are taken from the center of the bin which contains the PDF peak in Fig \ref{fig:conv_resp_fig} (bins are 10 keV wide). Symmetric 1$\sigma$ uncertainties in excitation energy and cross section are listed in parenthesis.}
  \centering
  \begin{ruledtabular}
  \begin{tabular}{crcc}
%    \hline
%    \hline
  \multicolumn{1}{c}{\text{Excitation}} &  \multicolumn{1}{c}{\text{Factor}} & \multicolumn{1}{c}{\text{Energy}} & \multicolumn{1}{c}{\text{Cross}} \\
  \multicolumn{1}{c}{\text{Energy}} &  \multicolumn{1}{c}{($\alpha$,$\gamma$)} & \multicolumn{1}{c}{\text{Bin Size}} & \multicolumn{1}{c}{\text{Section}} \\
  \multicolumn{1}{c}{\text{(MeV)}} &  \multicolumn{1}{c}{\text{to}} & \multicolumn{1}{c}{\text{FWHM}} & \multicolumn{1}{c}{\text{$^{15}$N($\alpha$,$\gamma$)$^{19}$F}}  \\
             &  \multicolumn{1}{c}{($\gamma$,$\alpha$)} & \multicolumn{1}{c}{\text{(MeV)}}      & \multicolumn{1}{c}{\text{(b)}} \\
  \hline
  4.755 (0.014) & 96.5 & [4.66, 4.79] & \multicolumn{1}{l}{1.5 (1.4) $\times 10^{-10}$} \\
  4.845 (0.014) & 104.2 & [4.74, 4.90] & \multicolumn{1}{l}{9.0 (8.6) $\times 10^{-10}$} \\
  4.945 (0.014) & 112.1 & [4.83, 5.01] & \multicolumn{1}{l}{1.1 (1.0) $\times 10^{-9}$} \\
  5.035 (0.014) & 118.5 & [4.91, 5.10] & \multicolumn{1}{l}{3.1 (2.6) $\times 10^{-9}$} \\
  5.135 (0.014) & 125.1 & [5.01, 5.20] & \multicolumn{1}{l}{3.9 (1.9) $\times 10^{-9}$} \\
  5.345 (0.015) & 137.1 & [5.33, 5.35] & \multicolumn{1}{l}{3.6 (0.8) $\times 10^{-6}$} \\
  5.345 (0.015) & 137.1 & [5.33, 5.35] & \multicolumn{1}{l}{1.3 (0.5) $\times 10^{-5}$} \\
  \end{tabular}
  \end{ruledtabular}
\end{table}
\endgroup

In comparison to the previous bubble chamber measurements at HI$\gamma$S from Ugalde et al. \cite{Ugalde} in 2013 (green crosses in Fig. 
\ref{fig:cs_fig}) the higher luminosity at Jefferson Lab have allowed us to push measurements to lower energies. At the lowest 
energy measurement, using an electron beam with $T_e$ = 4.813 MeV, the target luminosity is 3.5$\times$10$^{31}$ cm$^{-2}$
sec$^{-1}$. The wider horizontal error bars seen in the Ugalde et al. data arise from a different definition of the energy uncertainty used in this experiment. The 
Ugalde et al. data uses the width of the $\gamma$-ray distribution produced by the Compton scattering to determine the energy uncertainty of their cross section 
measurements. This gives energy uncertainties closer to the FWHM or energy bin size we 
determined from the PDFs shown in Fig. \ref{fig:conv_resp_fig}. A more complete comparison between the Ugalde et al. data and the cross section 
from this work would require an analysis using an $R$-Matrix calculation. An $R$-Matrix analysis of the cross section data from the Wilmes 
et al. \cite{Wilmes2}, Ugalde et al. \cite{Ugalde} and the Jefferson Lab 2018 results shown here, will be performed in an upcoming work.

\section {Summary and Future Plans} \label{sec:summary}

Compared to our earlier experiment \cite{Ugalde}, the cross section limits of the  
$^{15}$N($\alpha,\gamma$)$^{19}$F reaction have been pushed down by over one order of 
magnitude reaching cross sections of $\approx$ 100 pb. 
Although we are not yet in the single picobarn range, further improvements of this technique are possible. Due to the high luminosity
of this technique achieved through the use of thick targets ($\approx$ 10 g/cm$^2$) and a higher
phase-space factor, the present limits of the cross sections arise mainly from background
events of reactions in the wall of the glass vessel and the mineral oil. These events can be further reduced
by improving the neutron shielding, by eliminating the $^{10}$B content in the glass (through the use of pure Si-quartz glass), and by replacing the mineral oil with a hydrogen-free fluid.
As shown in Ref. \cite{DiGiovine} pictures taken by two cameras allows us to generate a 3-dimensional view of the bubbles and, thus, a method to eliminate the events outside of the fiducial region. Space limitations did not allow us to install two cameras in the present setup. We have now designed a compact stereo camera setup that will allow us to determine the 3D location of the bubble even under the space restrictions of the existing pressure vessel. With these improvements, cross section measurements of ($\alpha,\gamma$) reactions
in the picobarn range should be accessible.

In the future, studies of other helium-induced nuclear reactions of importance in many astrophysical scenarios should also become possible. The method used
in this work is sufficiently general to enable measurement involving other superheated fluids, or compounds soluble in a given superheated fluid which, contain 
monoisotopic or otherwise adequately depleted/enriched nuclei of interest. A bubble chamber with a liquid containing a magnesium solution could be used to study the 
photodisintegration of $^{26}$Mg to determine 
the rate of the $^{22}$Ne($\alpha$,$\gamma$)$^{26}$Mg neutron poison reaction \cite{22Ne_a_g_26Mg}. 
The weak component of the s-process is responsible for the production of nuclei with
60$\le$A$\le$90 in massive stars. It requires a neutron density of $\approx$ 1$\times$10$^{12}$ cm$^{-3}$,
which is provided mainly by the $^{22}$Ne($\alpha,n$)$^{25}$Mg reaction. The number of neutrons produced depends not only on its cross section but also on the rate of 
the $^{22}$Ne($\alpha$,$\gamma$)$^{26}$Mg reaction ---an alternate competing process that ``poisons'' 
the production of neutrons. Both rates are uncertain at temperatures relevant to the weak component of the s process.

Finally, the  $^{12}$C($\alpha$,$\gamma$)$^{16}$O reaction has been studied in normal and inverse 
kinematics for more that 50 years and cross section limits in the $\sim$pb range have 
been obtained \cite{deBoer}. For measurements of the photodissociation reaction
$^{16}$O($\gamma$,$\alpha$)$^{12}$C using a bubble chamber oxygen-containing superheated
liquids of H$_2$O, CO$_2$ and N$_2$O have been tested so far \cite{DiGiovine0}. 
The main difficulty for this reaction originates from the competing $^{17}$O($\gamma$,$\alpha$)$^{13}$C
and $^{18}$O($\gamma$,$\alpha$)$^{14}$C reactions with $Q$-values of -6.357 MeV and -6.227 MeV,
respectively, which are smaller than the one for photodissociation of $^{16}$O ($Q$ = -7.162 MeV).
Thus, this measurement requires the use of highly-enriched oxygen, in order to eliminate the 
contributions from the $^{17,18}$O isotopes. Measurements of these background reactions are
planned in the near future.

\section{Acknowledgements}

The authors would like to thank the accelerator staff at the Thomas Jefferson National Accelerator Facility
for their support. We would also like to thank Maurizio Ungaro for his help with the GEANT4 simulations as well as Peter Mohr and Alan Robinson 
for helpful discussions. The authors are also grateful for the helpful comments and suggestions of an anonymous reviewer which improved the quality of the paper. This
work was supported by the US Department of Energy, Office of Nuclear Physics, under Contract 
No. DE-AC02-06CH11357 and DE-AC05-06OR23177. This work was also supported by the U.S. National Science Foundation under Grant No.
2110898.  

\bibliographystyle{apsrev4-2}
\bibliography{ref.bib}

% Appendices =================================================================

\appendix
\section{Determination of electron beam parameters} \label{app:1}
\begin{figure}[htp]
        \centering
        \includegraphics[width=.4\linewidth]{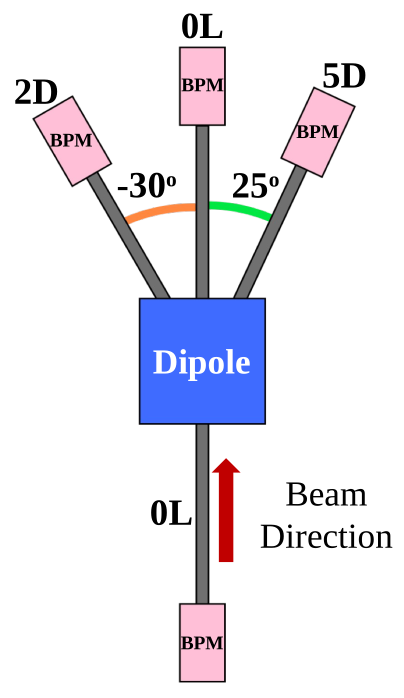}
        \caption{The 0L and spectrometer lines are co-planar in the horizontal plane.  The 2D and 5D lines form -$30^{\circ}$ and $25^{\circ}$ angles, respectively, with the straight ahead 0L line. The bubble chamber was installed on the 5D line.  Setting and measuring the electron beam characteristics for the experiment used the 0L, 2D, and 5D lines.}
        \label{fig:CartoonDumpLayout}
\end{figure}

The bubble chamber experiment ran in the injector portion of Jefferson Lab's Continuous Electron Beam Accelertor Facility (CEBAF) accelerator  tunnel. The quarter cryomodule, a superconducting radio frequency element, sets the beam momentum for the bubble chamber experiment. Downstream of the quarter cryomodule is a beam transport section with three beamlines served by a common dipole (see Fig.~\ref{fig:CartoonDumpLayout}): a straight ahead line (0L) to deliver beam to the next stage of acceleration before the beam is merged into the main CEBAF accelerator and two spectrometer dump lines (2D and 5D).  The bubble chamber equipment was installed in the 5D line.  The reported beam characteristics (momentum, momentum spread, beam size, and beam position and angle) at the bubble chamber's radiator rely on data recorded during the experiment from beam diagnostics installed in all three beam transport sections and simulation.

\subsection{Pre-experiment measurements}

\subsubsection{Earth's magnetic field}
All lines in Fig.~\ref{fig:CartoonDumpLayout} were shielded wherever possible with \text{$\mu$}-metal and low-carbon steel to reduce the background magnetic field around the beamlines.  Hall probe measurements above and below the unshielded segments of the lines characterized the background magnetic field for use in beam simulations.

\subsubsection{Reference orbit}
Beam position monitors (BPMs) measured the centroid of the beam, and their position readbacks provide a record of the beam orbit.  Quadrupole magnets in each beamline control the beam size, focusing the beam in one transverse plane while defocusing it in the other plane.  For the quadrupoles to be most effective and for beam quality reproducibility, the beam should traverse through the magnetic center of the quadrupoles.  Establishing a record of the beam orbit where the beam is centered in the quadrupoles provides a reference orbit for each setup.  Having a reproducible orbit is important for setting and measuring the beam momenta for the experiment.

The first step for providing orbit reproducibility took place prior to beam delivery and was an "as-found" mechanical survey of the positions of the quadrupole magnets and BPMs to establish their relative positions.  During machine setup, the second step was to define the reference beam orbit.  After threading the beam through the beamline, the beam was centered in each quadrupole and the beam position in the neighboring BPM recorded.  Centering in a quadrupole involves repetitively cycling the quadrupole magnet setpoint over a subset of its full range around its operational or design setpoint. In addition to steering the beam with upstream small dipole magnets known as correctors or steering magnets until beam motion is no longer present on a downstream diagnostic like a BPM (or visible on a phosphorescent view screen known as a beam viewer).  A record of the beam positions for the beam orbit centered in the quadrupoles provided a reference for each new momentum setup and data for error analysis.

\subsection{Setting the momentum}

For each requested momentum value, the dipole was set to the corresponding calculated magnetic field, and the dipole cycled twice through its hysteresis curve to ensure dipole setting reproducibility.  The cavities in the quarter cryomodule were operated on-crest providing maximum energy gain from each cavity, and the gradient setpoints of the two cavities were adjusted to set the momentum of the beam to match the dipole setting for the desired beam momentum in the 5D line.  BPM position readbacks in the 5D line giving an average zero-position across the line was determined when the momentum matched the dipole setting.  A high precision Hall probe measured the field in the center of the dipole. This was used as a reference in the measurement process.

\subsection{Momentum determination}

\begin{figure}[htp]
        \centering
        \includegraphics[width=0.9\linewidth]{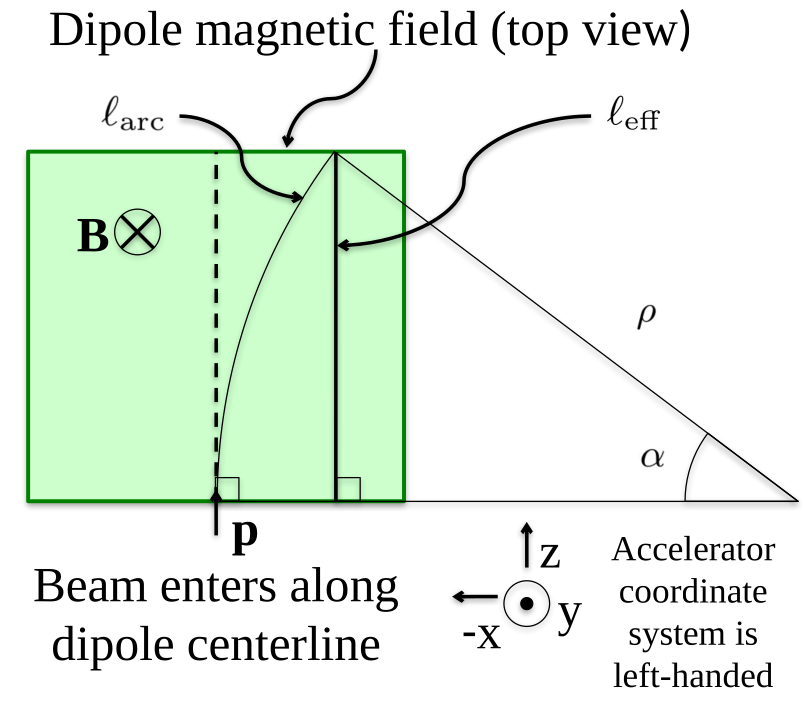}
        \caption{Considering the dipole from Fig.~\ref{fig:CartoonDumpLayout}, the electron beam traverses a circular arc, $\ell_{\text{arc}}=\rho \alpha$ ($\rho$ is known as the radius of curvature), through the dipole field, $\mathbf{B}$.  A convenient way to model a dipole is to use an ideal uniform dipole field, $B$, of finite extent, known as the effective length, $\ell_{\text{eff}}=\text{BDL}/B$ (=12.88 cm for our dipole). From the geometry, $\ell_{\text{eff}}=\rho\sin\alpha$, and the relation between the curved orbit and the effective length of the dipole is then $\ell_{\text{arc}}/\ell_{\text{eff}}=\alpha/\sin\alpha$.}
        \label{fig:CartoonDipoleDerivation}
\end{figure}

The two main ideas behind the momentum measurement for the experiment are that the dipole setting is a function of the beam momentum and that sending the beam to the 2D and 5D lines provides two different dipole settings for the same input beam conditions.  In general for beam in a dipole field (as illustrated in Fig.~\ref{fig:CartoonDipoleDerivation}), $p=qB\rho$ where $q$ is the electron charge, $B$ is the magnetic field, and $\rho$ is the radius of curvature. From the geometry, $p=qB\ell_{\text{arc}}/\alpha$, where $\ell_{\text{arc}}$ and $\alpha$ are the arc length and the angle of the orbit through the dipole field, respectively.  For the same input conditions to the 2D and 5D lines, $p=p_{2\text{D}}=p_{5\text{D}}$, and thus $\alpha_{2\text{D}}/\alpha_{5\text{D}}=B_{2\text{D}}\ell_{\text{arc}_{2\text{D}}}/(B_{5\text{D}}\ell_{\text{arc}_{5\text{D}}})$.  This relationship can be used to reconstruct the incoming momentum.  

The momentum measurement used all three beamlines. First, all quadrupole magnets in the 0L line downstream of the dipole and all magnets in the 2D and 5D lines were set to zero BDL, where   ${\text{BDL}}=\int \mathbf{B} \cdot \hat{\mathbf n} \times d\boldsymbol{\ell}$, and cycled through hysteresis. To send beam to the 0L line, the dipole was set to zero BDL. Next, the dipole setting was adjusted to send beam to the 2D line until the horizontal position at the end of the line was zero. Finally, the dipole setting was further adjusted to direct the beam to the 5D line until the horizontal position of the last BPM also read zero. The dipole set points use the $B\ell_{\text{eff}}$ model described in Fig.~\ref{fig:CartoonDipoleDerivation}. These set points are referred to as $\text{BDL}$s.  The recorded dipole settings for the 2D and 5D lines are summarized in Table~\ref{tab:beam_momenta}.

Rewriting the angle and field relationship in terms of $B\ell_{\text{eff}}$ and accounting for the difference between $\ell_{\text{arc}}$ and $\ell_{\text{eff}}$ gives
\begin{equation}
        \frac{\sin\alpha_{2\text{D}}}{\sin\alpha_{5\text{D}}} = \frac{[B\ell_{\text{eff}}]_{2\text{D}}}{[B\ell_{\text{eff}}]_{5\text{D}}}=\frac{[\text{BDL}]_{2\text{D}}}{[\text{BDL}]_{5\text{D}}}. \label{eq:alpha_bdl}
\end{equation}
Equation~\ref{eq:alpha_bdl} can be used to explore different sources of uncertainties. For example, the electron beam may have entered the dipole at an angle different from 90$^\circ$, which is a common source of angle uncertainty between the two lines. Another example would be if the angles of the two lines are slightly different from what is shown in Fig.~\ref{fig:CartoonDumpLayout}. A third example relates to the case where there is a common background magnetic field. All possible variations were examined and the results are presented below.

\begin{figure}[htp]
        \centering
        \includegraphics[width=1.0\linewidth]{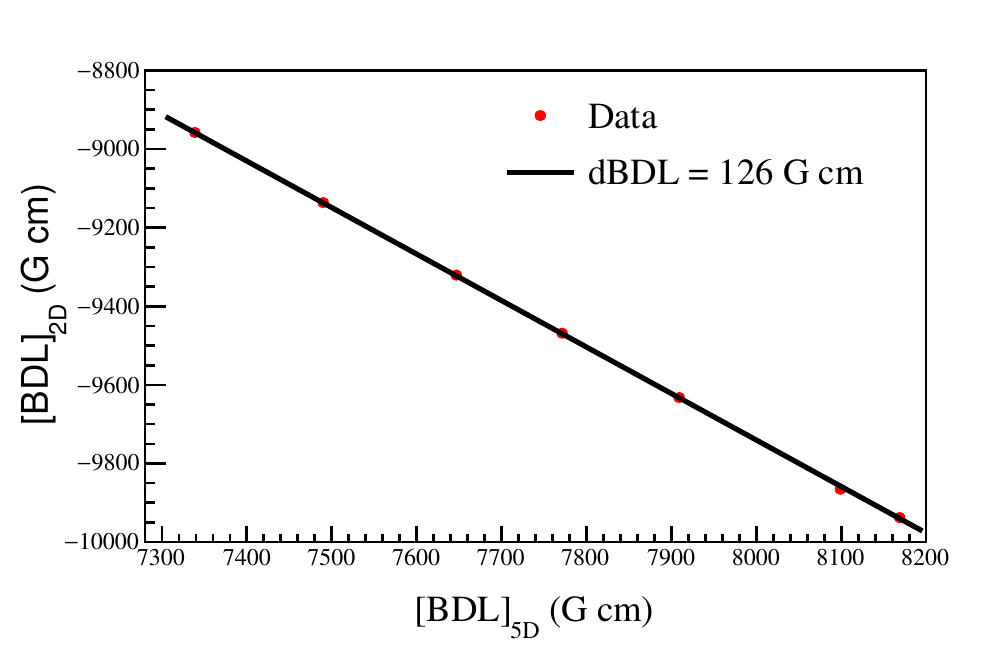}
        \caption{Dipole settings $[\text{BDL}]_{2\text{D}}$ versus $[\text{BDL}]_{5\text{D}}$ and linear least-squares fit of Eq.~\ref{eq:bdl_only_fit}.}
        \label{fig:BDLfits}
\end{figure}

As an example, assuming a common source of the BDL uncertainty ($\text{dBDL}$), Eq.~\ref{eq:alpha_bdl} can be written as
\begin{align}
        \frac{\sin\alpha_{2\text{D}}}{\sin\alpha_{5\text{D}}} & =\frac{[\text{BDL}]_{2\text{D}}+\text{dBDL}}{[\text{BDL}]_{5\text{D}}+\text{dBDL}}, \, \text{or}& \nonumber \\
        [\text{BDL}]_{2\text{D}} & =\frac{\sin\alpha_{2\text{D}}}{\sin\alpha_{5\text{D}}}[\text{BDL}]_{5\text{D}}+\left(\frac{\sin\alpha_{2\text{D}}}{\sin\alpha_{5\text{D}}}-1\right)\text{dBDL}. & \label{eq:bdl_only_fit}
\end{align}
Figure~\ref{fig:BDLfits} shows linear least-squares fit of Eq.~\ref{eq:bdl_only_fit}. This fit provided an updated $[\text{BDL}]$, which was used to calculate the corrected momentum. The final measured momentum is the average over all possible variations of $\alpha$ and $[\text{BDL}]$ resulting in good fits with root-mean-square (rms) of 0.007 MeV/c. Table~\ref{tab:beam_momenta} lists the measured beam momenta for the experiment.  The associated uncertainties for this measurement are summarized in Table~\ref{tab:beam_momenta_errors}.

\begin{table}[h]
  \caption{\label{tab:beam_momenta}Dipole settings to send the electron beam to the 2D and 5D lines and the corresponding measured beam momenta}
  \centering
  \begin{ruledtabular}
  \begin{tabular}{lccc}
    % \hline
    % \hline
                 &  {2D line} & {5D line} &               \\
    {Requested}   & {dipole}  & {dipole}  & {Measured} \\
    {$p$}       & {setting} & {setting} & {$p$}        \\
    {(MeV/c)} &  {(G cm)}  & {(G cm)}  & {(MeV/c)}  \\
    \hline
    5.24         & -8957.7    & 7338.9     & 5.299         \\
    5.34         & -9136.0    & 7490.0     & 5.406         \\
    5.44         & -9320.7    & 7646.8     & 5.517         \\
    5.54         & -9468.5    & 7771.4     & 5.605         \\
    5.64         & -9632.3    & 7909.2     & 5.703         \\
    5.74         & -9865.5    & 8099.0     & 5.840         \\
    5.84         & -9937.6    & 8168.8     & 5.887         \\
    % \hline
    % \hline
  \end{tabular}
  \end{ruledtabular}
\end{table}

\begin{table}[h]
  \caption{\label{tab:beam_momenta_errors}Uncertainties of the measurements of the electron beam momenta}
  \centering
  \begin{ruledtabular}
  \begin{tabular}{lc}
    % \hline
    % \hline
                                               & {Value} \\
    {Contribution}                             & {(\%)}  \\
    \hline
    Dipole power supply calibration (2 mA)               & 0.06       \\
    Dipole power supply regulation (1.5 mA)              & 0.04       \\
    Dipole field map offset (7 G cm) & 0.08       \\
    Dipole model                     & 0.10       \\
    Tracking model (0.007 MeV/c)                  & 0.12       \\
    \hline
    Total                                         & 0.19       \\
    % \hline
    % \hline
  \end{tabular}
  \end{ruledtabular}
\end{table}

\subsection{Momentum spread and beam size at the radiator}

In addition to magnets and BPMs, the three beamlines are instrumented with wire scanners for measuring the beam size. Using an \texttt{ELEGANT}~\cite{aps_elegant} model for the optics in the individual lines and measurements from the wire scanners, simulations provided the momentum spread of the beam and the beam size at the radiator. These results are presented in Table \ref{tab:mom_spread_beam_sizes}.

\begin{table}[h]
  \caption{\label{tab:mom_spread_beam_sizes}Measured momentum spread (d$p/p$) of the electron beam and the horizontal and vertical beam sizes at the wire scanner in the 5D line and the extrapolated beam sizes at the radiator. Note that beam size variation was unintended and limited experimental time precluded investigation.}
  \centering
  \begin{ruledtabular}
  \begin{tabular}{l|l|ll|ll}
    % \hline
    % \hline
                 &             & \multicolumn{2}{c|}{Horizontal} &  \multicolumn{2}{c}{Vertical} \\
                 &             & {Wire} &  {Radiator} & {Wire} &  {Radiator}  \\
    {Measured}&             & {rms}     & {rms}  & {rms}     & {rms}  \\
    {$p$}       &  {d$p/p$}  & {size}    & {size} & {size}    & {size} \\
    {(MeV/c)} & {($\times 10^{-3}$)} & {(mm)}    & {(mm)} & {(mm)}    & {(mm)} \\
    \hline
    5.299       & 1.76        & 1.31         & 1.70      &  0.70        & 0.57 \\
    5.406       & 0.311\footnote{We were generally unable to maintain this low d$p/p$.}      & 0.75         & 0.78      &  0.99        & 1.22   \\
    5.517       & 1.27        & 1.11         & 1.51      &  2.30        & 2.79   \\
    5.605       & 1.17        & 0.15         & 0.41      &  1.01        & 1.26   \\
    5.703       & 1.28        & 0.91         & 1.02      &  1.14        & 1.18   \\
    5.840       & 1.50        & 0.57         & 0.51      &  0.45        & 0.53  \\
    5.887       & 1.88        & 1.34         & 1.62      &   0.41       & 0.48 \\
    % \hline
    % \hline
  \end{tabular}
  \end{ruledtabular}
\end{table}
\subsection{Beam position and angle at the radiator}

With a model of the 5D beamline elements between the dipole and the radiator (3 corrector pairs, 2 quadrupoles, and 2 BPMs) including the background magnetic field, General Particle Tracer (\texttt{GPT}) \cite{pulsar_physics_gpt} simulations provided estimates of the position and angle of the beam on the radiator (Table~\ref{tab:positions_angles}).  The simulations used the measured beam positions from the BPMs and the setpoints for the corrector and quadrupole magnets to determine the likely beam position and angle at the radiator. The number of $\gamma$-rays reaching the bubble chamber was significantly reduced for the second electron beam setup because the vertical position was far off center. For the first beam setup, the effect of the horizontal position was much less significant. All beam properties were included in GEANT4 simulations to correctly generate the $\gamma$-ray spectra.  

\begin{table}[h]
  \caption{\label{tab:positions_angles}Beam positions and angles at the radiator (in a right-handed coordinate system)}
  \centering
  \begin{ruledtabular}
  \begin{tabular}{lccccc}
    % \hline
    % \hline
    {Measured} & \multicolumn{2}{c}{Horizontal} & \multicolumn{2}{c}{Vertical} \\
    {$p$}        & {position}      & {angle}   & {position}    & {angle} \\
    {(MeV/c)}  & {(mm)}     & {(mrad)}       & {(mm)}   & {(mrad)}     \\
    \hline
    5.299         & ~2.26           & -0.64            & -1.15         & -1.06         \\
    5.406         & ~0.99           & -1.90            & -5.24         & -3.42         \\
    5.517         & -0.29          & -1.63            &  ~0.10         & -0.38        \\
    5.605         & -0.78          & -3.67            & -1.17         & -1.17         \\
    5.703         & ~0.45           & -2.36            & ~0.23          & -0.39         \\
    5.840         & ~1.02           & -2.30            & -0.46          & -0.66         \\
    5.887         & ~0.95           & -3.58            &  ~0.86         &  ~4.02         \\
    % \hline
    % \hline
  \end{tabular}
  \end{ruledtabular}
\end{table}

\section{Description of yield model} \label{app:2}

The yield model (dashed black line in Fig. \ref{fig:yield_fig}) is the sum of two components. The
first is the convolution of a model of the cross section for the $^{19}$F($\gamma$,$\alpha$)$^{15}$N
reaction with the bremsstrahlung spectra calculated from GEANT4 (detailed in Sec. \ref{subsec:GEANT4_sim}). The cross section was modeled with Breit-Wigner curves, $\sigma_{BW}$(E), given by 
\begin{equation}
    \sigma_{BW}(E) = \pi\lambdabar^2\frac{\omega\Gamma_{\alpha}\Gamma_{\gamma}}{(E-E_R)^2 + (\Gamma/2)^2}
    \label{eq:BW}
\end{equation}
with $\lambdabar$ the de Broglie wavelength, $\omega$ the statistical factor from angular momenta ($J,J_1,J_2$), partial $\alpha$-particle width $\Gamma_{\alpha}$, partial 
$\gamma$-ray width $\Gamma_{\gamma}$, total resonance width $\Gamma$ and resonance energy $E_R$. Over the energy ranges of this work, the cross section for 
$^{19}$F($\gamma$,$\alpha$)$^{15}$N is dominated by the strong resonance at $E_R$ = 5.337 MeV with $J^{\pi}$ = 1/2$^{+}$. The cross section is an
incoherent sum of four single level Breit-Wigner curves representing the resonances at $E_R$ = 5.337, 5.535, 5.938 and 6.088 MeV, with widths from \cite{Wilmes2}. 

The second component of the model accounts for the background from the $^2$H($\gamma,n$)$^1$H reaction (described in Sec. \ref{sec:detection_and_backgrounds}), using the same convolution technique for the previously discussed $^{19}$F($\gamma$,$\alpha$)$^{15}$N reaction. The cross section for $^2$H($\gamma,n$)$^1$H (see Fig. \ref{fig:bkg_cs}) was convoluted with the GEANT4 bremsstrahlung spectra. 
From both the shape of the $^2$H($\gamma,n$)$^1$H cross section (being relatively flat above 3.0~MeV) and the shape of the bremsstralung spectrum (for example, the $T_e$ = 4.813 MeV spectra shown in Fig. \ref{fig:geant4_brem_spectra}), the convolution resembles a straight line in a semi-log plot. This line was modeled by $y = k10^{m \times x}$, with a slope $m$ and $y$-intercept $k$.  In this context, the $y$-intercept can be seen as a normalization constant. Here we determine the value of $k$ from the experimentally measured yield for an electron beam at $T_e$ = 4.0 MeV, that is $k$ satisfies the condition that 
\begin{equation}
    Y_{exp}(4.0\, \text{MeV}) =  k10^{m \times 4.0}.
    \label{eq:k-def}
\end{equation}
From a linear regression over the range of electron energy from 4.6 - 5.1 MeV (where the deuterium contribution to the yield is the largest) and a Monte Carlo calculation over several samples, we determined $m$ = 9.19$\times$10$^{-1}$ $\pm$ 2.16$\times$10$^{-2}$ and $k$ = 3.21$\times$10$^{-8}$ $\pm$ 8.24$\times$10$^{-9}$. 

The sum of these two contributions (yield resulting from $^{19}$F photodisintegration and the yield resulting from neutrons produced via $^2$H($\gamma,n$)$^1$H) is the theoretical yield. A Monte Carlo calculation sampling over electron energies from 4.0 to 5.5 MeV was used to produce the curve bound by the dashed red lines seen in Fig \ref{fig:yield_fig}.

\end{document}